%% file: draft.tex
\documentclass[12pt,a4paper]{article}%
\pdfoutput=1
\usepackage{jheppub}
\usepackage{amsmath,amssymb,amsfonts,wasysym}
\usepackage{mathrsfs}
\usepackage[bf,footnotesize]{caption2}
\usepackage{bbm}
\usepackage{ulem}
\usepackage{url} 
\usepackage{latexsym,epsfig,amssymb,amsmath}
\usepackage{verbatim}

\def\({\left(}
\def\){\right)}

\def\be{\begin{equation}}
\def\ee{\end{equation}}
\def\bry{\begin{array}}
\def\ery{\end{array}}
\def\bes{\begin{subequations}}
\def\ees{\end{subequations}}
\def\bit{\begin{itemize}}
\def\eit{\end{itemize}}
\def\ben{\begin{enumerate}}
\def\een{\end{enumerate}}

 \def\be{\begin{equation}} \def\ee{\end{equation}}
\def\bea{\begin{eqnarray}} \def\eea{\end{eqnarray}}

\newcommand{\non}{\nonumber}
\newcommand{\Tr}{{\rm Tr}}

\newcommand{\GeV}{{\rm\ GeV}}

\newcommand{\TeV}{{\rm\ TeV}}
\newcommand{\fb}{{\rm\ fb}}

\newcommand{\Dsl}{D\llap{/\kern+1.5pt}}
\newcommand{\MET}{E\llap{/\kern1.5pt}_T}

\def\Zuu{Z_{u_i u_j}}

\title{
\begin{center}
Signs of Tops from Highly Mixed Stops
\end{center}
}
\author[a]{Mihailo Backovi\'{c}}
\author[b]{Alberto Mariotti}
\author[c]{Michael Spannowsky}
\affiliation[a]{
Center for Cosmology, Particle Physics and Phenomenology - CP3, \\
Universite Catholique de Louvain, Louvain-la-neuve, Belgium}
\affiliation[b]{Theoretische Natuurkunde and IIHE/ELEM, 
 Vrije Universiteit Brussel,\\
 and International Solvay Institutes, Pleinlaan 2, 
 B-1050 Brussels, Belgium}
\affiliation[c]{Institute for
  Particle Physics Phenomenology, Department of Physics,\\
  Durham University, Durham DH1 3LE, United Kingdom}
\emailAdd{mihailo.backovic@uclouvain.be}
\emailAdd{alberto.mariotti@vub.ac.be}
\emailAdd{michael.spannowsky@durham.ac.uk} 
\proceeding{\vspace{-1.65cm}\flushright{\small IPPP/15/07\\ DCPT/15/14\\}}
\abstract{
Supersymmetric extensions of the Standard Model with highly mixed squark flavours beyond minimal flavour violation provide interesting scenarios of new physics,  which have so far received limited attention.
 We propose a calculable realization of such scenarios in models of gauge mediation augmented with an extra interaction between the messengers and the up type squark.
We compute the supersymmetric spectrum and analyze the flavour physics constraints on such models. 
In a simplified model approach, we show that scenarios with maximal squark flavour mixing result in interesting phenomenological signatures at the LHC. 
We show that the model can be probed up to masses of $m_{\tilde{u}} \lesssim 950 \GeV$ in the single-top event topology at LHC14 with as little as 300 fb$^{-1}$. The
 most distinctive signature of highly mixed scenarios, the same sign positive charge di-top, can be also probed to comparable squark masses at high luminosity LHC14.
}

\begin{document} 
\maketitle
\setcounter{page}{2}


\section{Introduction}

Recent results from the Large Hadron Collider (LHC) impose stringent limits on the scale of supersymmetry and considerably constrain the 
minimal supersymmetric extension of the Standard Model (MSSM). Yet, the spectrum of  non-standard supersymmetric (SUSY) models has not been studied in full detail, while the current data constraints on non-standard SUSY are weaker \cite{Bozzi:2005sy,Dittmaier:2007uw,Fuks:2008ab,Hurth:2009ke,Crivellin:2010gw,Blanke:2013uia}. At the dawn of the LHC Run II at $\sqrt{s} =14$ TeV,  and with the prospects for  the future high luminosity LHC, it is important to revisit constraints imposed in studies of the MSSM parameter space, and to explore if some viable realizations have been overlooked and/or could be accessed through new types of collider signatures.  

As an example of scenarios which were previously less studied, one could consider relaxing the minimal flavour violation (MFV) assumption on SUSY models. One interesting option for non-standard SUSY beyond MFV involves  considerations of mixing between squark flavours.  The immediate benefit of scenarios where squark mixing is allowed is that branching ratios of squark decays into final states predicted by standard SUSY models are lowered, hence weakening the experimental bounds on the stop mass. Current experimental results, particularly from flavour physics, give strong constraints on mixing between first and second generation squarks, while mixing between third and first/second generations remains relatively unconstrained. Several scenarios of SUSY models with squark mixing have recently been discussed in Refs.~\cite{Blanke:2013uia, Agrawal:2013kha, Kribs:2009zy, Plehn:2009it,Gedalia:2012pi}.

In this paper we investigate the possibility that the low energy part of the supersymmetric spectrum is characterized by a single (right handed) light squark, highly mixed in flavour between stop and up-squark or stop and scharm-squark.  We propose an explicit realization of such scenarios in a simple extension of gauge mediated models, augmented with an extra interaction between the messengers and the right up type quark superfield. We study the constraint imposed on such models by flavour observables and by the requirement of a viable Higgs mass.
A similar extension of the gauge mediation scheme has been recently studied in several different contexts
\cite{Chacko:2001km,Delgado:2007rz,Giudice:2007ca,DeSimone:2011va,Evans:2011bea,Shadmi:2011hs,
Abdullah:2012tq,Albaid:2012qk,Craig:2012xp,Kang:2012ra,Evans:2012hg,Evans:2013kxa,Calibbi:2013mka,Galon:2013jba,Byakti:2013ti,Calibbi:2014pza, Evans:2015swa}, 
for instance
with the purposes of addressing the $\mu-B_{\mu}$ problem, of raising the Higgs mass and/or investigating their possibly unusual flavor patterns.

We study in detail the reach of LHC14 for the case where the light squark is maximally mixed between top- and up-type squark.
We focus on the collider signatures which are peculiar to a scenario with maximal stop-sup mixing, i.e. single top and same sign di-top
production.
These signatures have already been investigated in several new physics scenarios.
The single top signatures have been studied for instance in \cite{Arhrib:2006sg,Aranda:2010cy,Andrea:2011ws,Agram:2013wda,Alvarez:2013jqa,Boucheneb:2014wza}.
Same sign tops have been studied in new physics models attempting to explain the forward backward asymmetry
\cite{Jung:2009jz,Gupta:2010wx,Degrande:2011rt,Cao:2011ew,Berger:2011ua,Atwood:2013xg,Goldouzian:2014nha}, or
in composite models \cite{Contino:2008hi,Mrazek:2009yu}.
In the context of supersymmetry, same-sign top final states can arise either in 
R-parity violating models
\cite{Durieux:2012gj,Durieux:2013uqa,Csaki:2013jza} or in
R-parity preserving theories through a gluino decay chain
\cite{Kraml:2005kb,Martin:2008aw}. However,  in the latter case we expect a roughly  equivalent number of same sign positive top pairs and
of same sign negative top pairs. 
Instead, in scenarios with large sup-stop mixing,  the same-sign top contribution at the LHC will be dominantly of positive charge, since it is obtained through an initial 
state of two up quarks.
In the context of R-parity preserving MSSM, the presence of the single top signature in association with the
same sign positive top signature 
would hence be a robust hint of large squark mixing.

We define a simplified model for LHC searches that consists only of the gluino (that we fix representatively at $2$ TeV), 
the maximally mixed sup-stop right-type squark and the neutralino (mostly Bino).
Considering benchmark points not yet excluded by LHC8, 
we show that the single top signal can be discovered at LHC14 with 300 fb$^{-1}$ up to $m_{\tilde u_1} \lesssim 950 \GeV$ in the aforementioned channels.
The more distinctive signature of sizeable sup-stop mixing, (i.e. same sign tops), leading eventually to same sign positive leptons plus missing energy,
can be probed at High Luminosity LHC14.
Our collider analysis represents a concrete proposal to test the mixing property of the light squark in the next LHC runs.

The paper is organized as follows: In section \ref{model_formulation} we present the model setup and we explore the relevant parameter space, identifying the phenomenologically viable regions and possible benchmark points.
On the basis of this analysis, in section \ref{simplified_model} we define a simplified model for collider studies, 
with a light and maximally mixed sup-stop squark (MMUT). In section \ref{LHC_section1} we discuss its main production modes at collider and
the constraint on the model from LHC8.
In section \ref{LHC_section2} we study in detail the LHC14 signals of the simplified model MMUT, i.e. processes with a single top quark and missing energy $\MET$ or the distinctive same sign positive top with $\MET$ signature. We show the reach of LHC14 for these final states on some representative benchmark points of the MMUT model.

\section{The model formulation} \label{model_formulation}

\subsection{Low Energy Constraints}

In the minimal supersymmetric extension of the Standard Model (MSSM), the supersymmetry breaking parameters (the soft terms) 
can be generic sources of flavour and CP violation. 
Hence, low energy observables put strong constraints on the structure of the soft terms and more generally on the mechanisms of supersymmetry breaking and its mediation to the MSSM.
In this context, gauge mediation (see e.g. \cite{Giudice:1998bp}) is one of the most interesting possibilities, since it provides a predictive and computable framework while accommodating for the constraints from flavour physics.
In this paper we consider an extension of gauge mediation involving extra interactions that break the flavour degeneracy in a controllable way, by
modifying the structure of the soft terms.
Our purpose is to provide a predictive framework for supersymmetric models with large mixing in the right squark up sector, and with one light
squark eigenstate. In particular we will focus on RR mixing between the first and the third generation.

The flavour bounds in the MSSM can be analysed in a model independent way by constraining the structure of the supersymmetry breaking parameters.
Here we briefly review the constraints on the up-type squark mass matrix 
that are relevant in our analysis, while for a more comprehensive review we refer the reader to \cite{Altmannshofer:2009ne}.

The up-type squark mass matrix in the superCKM basis can be written as
\be
\mathcal{M}= \tilde m^2 (\mathbb{I}_{6 \times 6}+ \delta_u) \qquad 
\delta_u=\left( 
\begin{array}{c c}
\delta_u^{LL} & \delta_u^{LR}\\
\delta_u^{RL} & \delta_u^{RR}
\end{array}
\right)
\ee
where the dimensionless $\delta_u$ parameterise the deviations from flavour alignment. 
We are interested in flavour mixing showing up in the right-right part of the up-type squark mass matrix, since they are less constrained 
by flavour physics \cite{Dittmaier:2007uw}.

The matrix $(\delta^{RR}_u)_{ij}$ is a $3 \times 3$ matrix which is determined by the soft mass of the up-type squark $(m_{\tilde u}^2)_{ij}$.
The most relevant constraint on the mass matrix is obtained through the measurement of the $D_0-\bar D_0$ mixing, 
which bounds the absolute value of the $(\delta^{RR}_u)_{12}$ to be smaller than 0.05 \cite{Altmannshofer:2009ne,Ciuchini:2007cw}.
Apart from this bound, low energy observables do not independently constrain $(\delta^{RR}_u)_{13}$ or $(\delta^{RR}_u)_{23}$.

Another possibly relevant constraint to keep into account for the 1-3 mixing is given by the neutron electric dipole moment (EDM). 
Even in the case of no CP phases in the supersymmetry breaking parameters, a small deviation of the soft terms from being proportional to the identity matrix
can induce some extra CP violating effect once we rotate into the CKM basis. The strict experimental bounds on the neutron EDM can lead to stringent constraints on the allowed mass spectra \cite{Dedes:2014asa}.
However, the neutron EDM processes also involve L-R mixing, hence they can be relevant only in the combined presence of large 1-3 RR mixing and (diagonal or off diagonal) large LR mixing.

In the following we provide a computable model which induces large off diagonal contribution only to the RR up-type squark mass matrix $(m_{\tilde u}^2)_{ij}$
in the framework of extension of gauge mediation, and which is compatible with flavour constraints.

\subsection{Extended Gauge Mediation}\label{themodel}

We proceed to define a class of models with large squark mixing in the context of gauge mediated supersymmetry breaking. 
Scenarios with large squark mixing can be accommodated by extensions of gauge mediated models augmented with extra superpotential interactions among the messengers and some superfield of the MSSM.
The typical realization of gauge mediation (for a review see \cite{Giudice:1998bp}) includes messenger fields charged under the SM gauge groups. It is then natural to
explore the possibility that the messengers could couple directly to SM matter fields. Extending gauge mediation with extra superpotential
interactions in turn induces extra contributions to the soft terms.

This class of models
has recently been studied in several papers 
\cite{Chacko:2001km,Delgado:2007rz,Giudice:2007ca,DeSimone:2011va,Evans:2011bea,Shadmi:2011hs,
Abdullah:2012tq,Albaid:2012qk,Craig:2012xp,Kang:2012ra,Evans:2012hg,Evans:2013kxa,Calibbi:2013mka,Galon:2013jba,Byakti:2013ti,Calibbi:2014pza}.
An interesting aspect of the above mentioned superpotential deformations is that they are not automatically diagonal in flavour space, and hence they can represent 
controllable sources of flavour violation, which are normally absent in standard gauge 
mediation.\footnote{Another interesting aspect of these deformations, which we do not exploit here, is that they could also induce large A-terms.} In addition, the set of possible extra interactions involving the messengers (in complete representation of unified gauge group) and the SM matter fields have been classified in \cite{Evans:2013kxa},
where also the complete formulas for the induced soft masses have been computed.

Here we focus on a specific superpotential interaction involving messenger fields and the right handed type squarks, since we 
aim for the possibility of inducing large mixing in the right up squark mass matrix. 
For this reason our model differs from the one considered in \cite{Evans:2013kxa},
since it has a generic flavor pattern.
Moreover, contrary to most models considered in the literature,
it also respects a discrete R symmetry. This motivates the absence of other MSSM-messenger couplings and avoids the problematic issue of generating large off-diagonal A-terms
which are generically strongly bounded by low energy observables. Large off-diagonal A-terms  could also lead to other problematic issues in the
evolution of the SUSY spectrum, as we will discuss in the following.

The model we consider consists of a pair of messengers in the $\bar 5 + 5$ that we denote $(\phi_1,\phi_2,\tilde \phi_1,\tilde \phi_2)$.
We assume that the component of the $\bar 5$ messengers with the same quantum numbers of right handed down quark interact with the up type quarks via
the superpotential couplings
\be
\label{deformation}
\delta W = \lambda \sum_{i=1}^{3}  c_i U_i   \phi_{1(\bar 5,D)}   \phi_{2(\bar 5,D)},
\ee
where the index $i$ runs here on the flavour index.
Hence the interaction is not diagonal in flavor, and can induce non-trivial flavor mixing, depending on the values of the vector $\vec c=\{c_1,c_2,c_3 \}$.
We take the normalization $\sqrt{c_1^2+c_2^2+c_3^2}=1$, while $\lambda$ sets the overall size of the deformation. 
We take all couplings to be real, in order to not introduce sources of CP violation.
The superpotential (\ref{deformation}) adds to the usual MSSM superpotential which
includes the Yukawa couplings and the $\mu$ term.

The supersymmetry breaking superpotential for the two pairs of $5 +\bar 5$ messengers is 
\be
\label{R-preserving}
W_{R\text{-pres}}= M (\phi_1 \tilde \phi_1+\phi_2 \tilde \phi_2) + Y (\phi_1 \tilde \phi_2 + \phi_2 \tilde \phi_1),
\ee 
with the SUSY breaking spurion $Y =\theta^2 F$.
The messenger scale $M$ sets the energy scale where the soft masses are generated.
Note that there is a residual $\mathbb{Z}_4 $ $R$-symmetry
under which the spurion $Y$ has charge $2$, with the other charges reported in Table~\ref{discreteR}.
This implies that gaugino masses and also $A$-terms cannot be generated by this sector and by the deformation $\lambda$.
\begin{table}[h]
\begin{center}
\begin{tabular}{c||c|c|c|c|c|c|c}
& $Y$ & $ \phi_1$ & $ \tilde \phi_1 $ & $ \phi_2$ & $ \tilde \phi_2 $ & $U,D,E$ & $Q,H_u,H_d,L$ \\
\hline
$\mathbb{Z}_4 $ R-symmetry & 2 & 2 & 0 & 0 & 2 & 0 & 1\\
\end{tabular}
\caption{\label{discreteR}$\mathbb{Z}_4 $ $R$-symmetry charge assignment.}
\end{center}
\end{table}

The discrete R-symmetry,
together with a messenger $Z_2$ symmetry under which $\phi_i$ and $\tilde \phi_i$ are odd,
implies that the deformation (\ref{deformation}) is the only one compatible with these discrete simmetries and with gauge invariance.
Indeed the $Z_2$ symmetry of the messengers, together with gauge invariance, would allow only for the extra coupling $Q \phi_{1(\bar 5,D)} \phi_{2,(\bar 5,L)}$, 
which is forbidden by the discrete R-symmetry in Table \ref{discreteR}.

The boundary conditions induced at the messenger scale by this SUSY breaking sector includes gauge mediated contributions to the
scalar masses and the contribution arising because of the new interaction $\lambda$.
The soft masses for this model are obtained in the Appendix \ref{appendixA}, here we report simply the results, defining $\Lambda=\frac{F}{M}$.
The gauge mediation contribution is 
\be
\label{gmsb1}
m_{\tilde f}^2= \frac{2}{(16 \pi^2)^2} \sum_r C_r^{\tilde f} g_r^4 \Big( 2  \Lambda^2 f_s(\frac{\Lambda}{M}) \Big)
\ee
where $f_s(x)$ is the usual minimal gauge mediation function for sfermions (see e.g.~\cite{Martin:1996zb}) $f_s(x)=1 +\frac{x^2}{36} + O(x^4)$.
The contribution to the soft masses induced by the $\lambda$ deformation are
\bea
&&
 m_{U_{i} U_{j}}^2= \frac{1}{256 \pi^4} c_{ij} \lambda^2 d_U  \left(  \lambda^2 d_{\phi}   -2    \sum_{r=1,3} C_r g_r^2 \right) \Lambda^2
-\frac{d_U}{48 \pi^2} c_{ij} \lambda^2  h(\frac{\Lambda}{M}) \frac{\Lambda^4}{M^2} \nonumber \\
&&
\label{def_thresold_R}
 m_{Q_{ij}}^2=-\frac{d_U \lambda^2}{256 \pi^4} (y^{\dagger}.c.y)_{ij} ~\Lambda^2   \\
&&
m_{H_u}^2=-\frac{3 d_U \lambda^2}{256 \pi^4}  \Tr (y^{\dagger}.c.y) ~\Lambda^2 \nonumber 
\eea
where we defined the matrix $c_{ij}=c_i c_j$, $d_U=2$ and $d_{\phi}=4$, $C_1=2/5,C_2=0,C_3=4$.
We give the complete expression for the one loop function $h(x)$ in  Appendix~\ref{appendixA}. It can be approximated by $h(x) = 1+\frac{4}{5} x^2 + O(x^4)$.

The up type right squark gets off diagonal contributions whose flavour structure is determined by the matrix $c_{ij}$,
consisting of a two loop contribution (which can be positive or negative depending on the value of $\lambda$) 
and a one-loop negative contribution, which can be relevant
for not too small values of $\frac{\Lambda}{M}$.
Besides the up-type squark, the rest of the soft masses contributions are determined by the Yukawa couplings $y_{ij}$.
In particular the contribution to $m_{Q_{ij}}$, that would have implied otherwise strong bounds from flavor observables, is
projected along the Yukawa couplings.

Moreover, the fact that the hidden sector respects a discrete $R$ symmetry implies that we have not generated any A-term.
This is a welcome feature in perspective of possible large off-diagonal contributions induced by the $\lambda$ 
deformation. Indeed, large off diagonal
A-terms would have raised two problematic issues. First, large off diagonal A-term involving the first generation can lead to large 
neutron EDM, above the current experimental bound, de facto excluding the model, as we previously mentioned.
Second,
large off diagonal A-terms, together with gaugino masses, induces corrections to fermion masses which can be incompatible with the actual value of the light quark masses.
On the other hand, the absence of A-terms implies that in order to get the correct Higgs mass we will have to consider at least one stop to be quite heavy.

In order to induce non vanishing gaugino masses and sizable scalar masses
we consider also in addition to the previous SUSY breaking sector another supersymmetry breaking sector inducing general gauge mediation contribution, with different gaugino and scalar SUSY breaking scales, respectively $\Lambda_G$ and $\Lambda_S$
\bea
&&
\label{gmsb2}
M_{g_r}=\frac{g_r^2}{16 \pi^2}  \Lambda_G \\
&&
m_{\tilde f}^2= \frac{2}{(16 \pi^2)^2} \sum_r C_r^{\tilde f} g_r^4 \Lambda_S^2. \nonumber
\eea
For simplicity we assume that such contribution is induced at the same messenger scale $M$ as above,
which hence sets the range of scales of the renormalisation group (RG) flow.

In summary,  the total contribution to the soft masses in the complete model is given by adding expressions (\ref{gmsb1},\ref{def_thresold_R},\ref{gmsb2}). \\

We explored the parameter space of this model and the implication for flavor observables.
We implemented the model in SARAH \cite{Staub:2013tta,Porod:2014xia} and generated the spectrum and compute the flavor observables using SPheno \cite{Porod:2011nf}.
We also computed the contribution to $D_0$ mixing, which is sensitive to the mixing between the first and second generation squarks,
and the neutron EDM, using the SUSYFLAVOR code \cite{Crivellin:2012jv}. 

As a large contribution to $D_0$ mixing arises through hadronic long-distance physics \cite{Golowich:2005pt, Golowich:2007ka}, which is plagued by large theoretical uncertainties, for the limit setting we only impose an upper bound, i.e. its measured central value, on the short-distance SUSY contributions.
The flavour observables that we checked, together with the bounds that we applied, were taken from \cite{flavour_CKMfitter}
and are summarized in Table \ref{Flavour_bounds}.

\begin{table}[t]
\begin{center}
\begin{tabular}{|c|c|c|c|}
\hline
Flavour Observable & Imposed limit & Source &  Tool  \\
\hline
$\frac{BR(B \to X_s \gamma)}{BR(B \to X_s \gamma)_{SM}}$ & [0.84, 1.16] & HFAG \cite{hfag_pdf} & SPheno  \\
\hline
$\frac{BR(B^0_d \to \mu \mu)}{BR(B^0_d \to \mu \mu)_{SM}}$ & [0.87, 1.08] & HFAG \cite{hfag_pdf} & SPheno  \\
\hline
$\frac{BR(B^0_s \to \mu \mu)}{BR(B^0_s \to \mu \mu)_{SM}}$ & [0.90, 1.06] & HFAG \cite{hfag_pdf} & SPheno  \\
\hline
$\frac{\Delta M_{B_s}}{(\Delta M_{B_s})_{SM}}$ & [0.90, 1.17] & HFAG \cite{hfag_pdf} & SPheno \\
\hline
$\frac{\Delta M_{B_d}}{(\Delta M_{B_d})_{SM}}$ & [0.85, 1.13] & HFAG \cite{hfag_pdf} & SPheno \\
\hline
$\frac{BR(D \to \mu \nu)}{BR(D \to \mu \nu)_{SM}}$ & [0.95, 1.01] & HFAG \cite{hfag_pdf} & SPheno  \\
\hline
$\frac{BR(D_s \to \mu \nu)}{BR(D_s \to \mu \nu)_{SM}}$ & [0.95, 1.03] & HFAG \cite{hfag_pdf} & SPheno  \\
\hline
$\frac{BR(K \to \mu \nu)}{BR(K \to \mu \nu)_{SM}}$ & [0.99, 1.01] & HFAG \cite{hfag_pdf} & SPheno  \\
\hline
$\frac{BR(B \to s \mu \mu)}{BR(B \to s \mu \mu)_{SM}}$ & [0.12, 1.87] & HFAG \cite{hfag_pdf} & SPheno  \\
\hline
$\frac{BR(B \to K \mu \mu)}{BR(B \to K \mu \mu)_{SM}}$ & [0.875, 1.125] & HFAG \cite{hfag_pdf} & SPheno  \\
\hline
$\frac{\epsilon_K}{(\epsilon_K)_{SM}}$ & [0.68, 1.34] & \cite{pdg} & SPheno \\
\hline
$\frac{\Delta M_K}{(\Delta M_K)_{SM}}$ & [0.997, 1.003] & \cite{pdg} & SPheno \\
\hline
$\Delta M_D$ &  $\leq 8.82 \times 10^{-15}$ GeV & \cite{Amhis:2014hma} & SUSYFLAVOR \\
\hline
Neutron EDM $|d_n|$ &  $\leq  2.9 \times 10^{-26}$ (e cm) & \cite{Baker:2006ts} & SUSYFLAVOR \\
\hline
\end{tabular}
\end{center}
\caption{\label{Flavour_bounds} Flavour bounds imposed during the scan on the model parameter space.}
\end{table}

\begin{figure}[t]
\begin{center}
\includegraphics[width=0.485\textwidth]{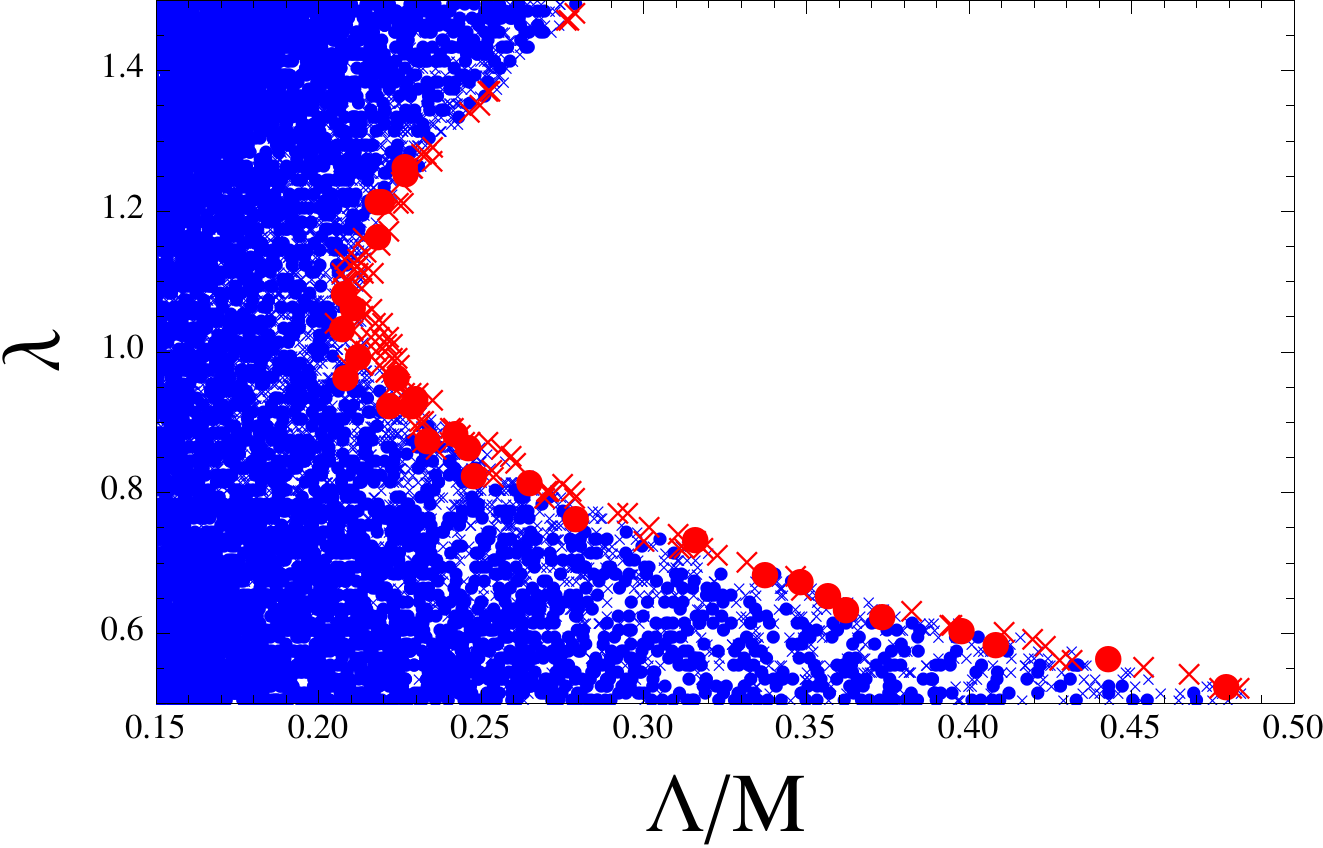}
~
\includegraphics[width=0.485\textwidth]{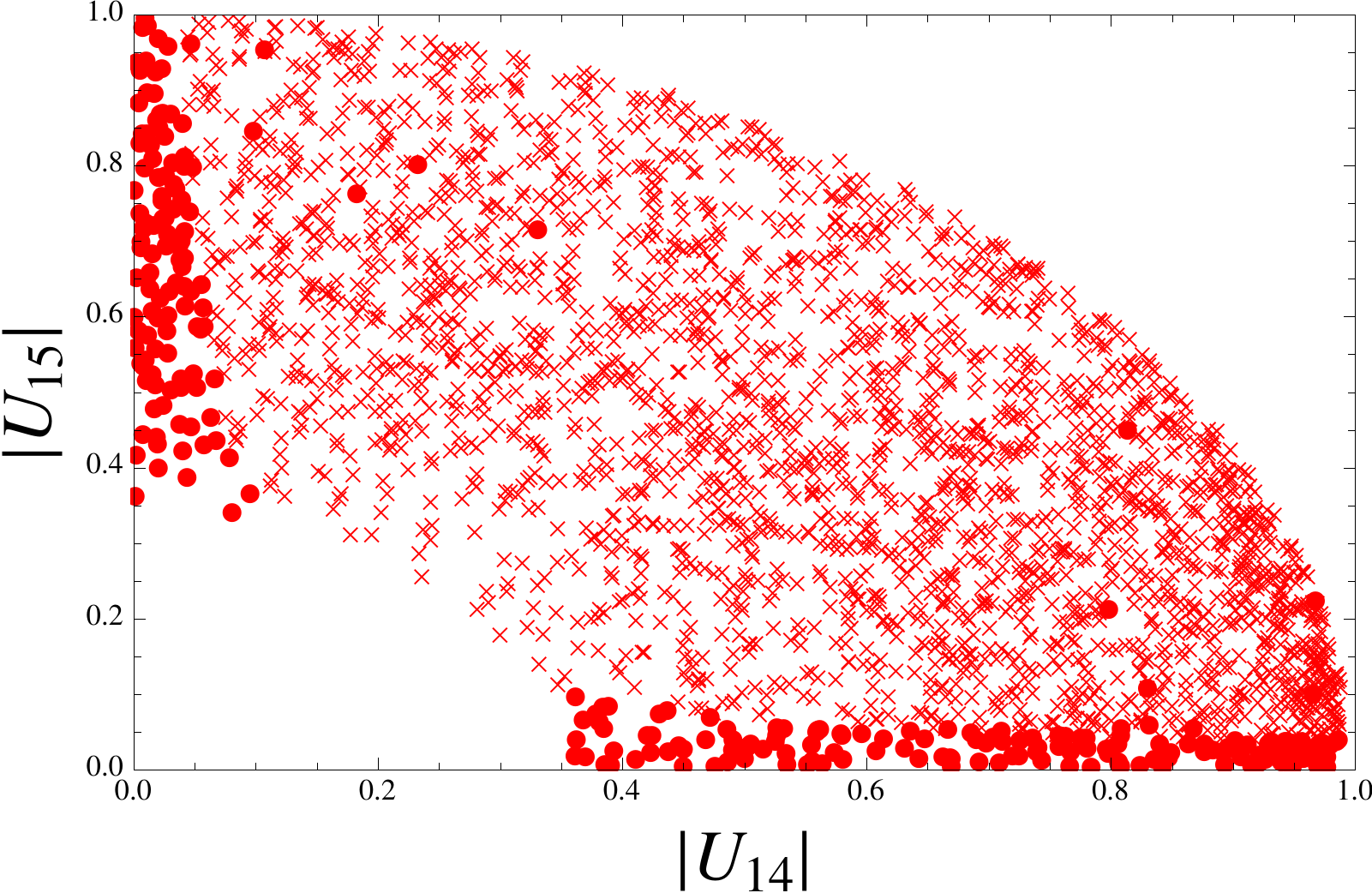}
\caption{\label{lambda_scan} Left: results of the scan on the parameter space of the model on the $(\lambda,\Lambda/M)$ plane. 
We fixed $\Lambda_S=1 \times 10^6$ GeV, 
$\Lambda_G=1.9 \times 10^5$ GeV (corresponding to gluino mass around $1.5$ TeV), $\Lambda=5 \times 10^5 $ GeV. We scanned over $\lambda,c_1,c_2,c_3$ and $M$. The red points have one squark with mass smaller than 1.5 TeV.
Right: we focus on a region with light squark: we fixed $\lambda=1$ and we restricted $M$ to be such that
 $\Lambda/M\simeq 0.2 $, varying freely on $(c_1,c_2,c_3)$; we show only points having one light squark with mass smaller than 1.2 TeV, and 
 we plot in the physical mixing angle plane $(|U_{14}|, |U_{15}|)$, where $U_{14}$ is the up component and $U_{15}$ is the charm component of the lightest squark (see eqn (\ref{mixing_conv})). All points displayed have $m_h = 125 \pm 2.5$ GeV.}
\end{center}
\end{figure}

In order to establish general low energy physics constraints on the model, we performed a scan by fixing the values of $\Lambda_G$, $\Lambda_S$ and $\Lambda$ and varying the messenger mass $M$, the deformation size $\lambda$, and the flavour direction $\vec c$ of the deformation. In the left plot of Fig. \ref{lambda_scan} we show the result in the $(\lambda,\Lambda/M)$ plane.
The points respecting flavour constraints are shown as circular dots, while the crosses are points violating flavour observables.
The red points are scenarios where the lightest up type squark is lighter than $1.5$ TeV.
The Higgs mass is correct (within the errors) on all the points shown in the plot, due to the large value of $\Lambda_S$.
Since we are marginalizing over the flavour direction, there are overlapping points which have the same
size of the deformation but which can or cannot satisfy the flavour bounds.
However the plot is useful in order to understand the effect of the deformation on the soft spectrum, independently on the direction $\vec c$,
as we explain now.

The main effect of the deformation is on the mass squared $m_{U_{ij}}^2$ (see eqn. (\ref{def_thresold_R})), 
which can be negative (or very small) 
at the messenger scale and hence positive but small at the EW scale, in the region of moderate $\lambda$ and large ratio $\Lambda/M$.
In particular the final spectrum is very sensitive to the value of the ratio $\Lambda/M$, because it determines 
how large is the one loop negative contribution to the soft mass
$m_{U_{ij}}^2$ 
at the messenger scale.
The allowed region in Fig. \ref{lambda_scan} is determined by the fact that for large values of $\Lambda/M$ the negative one loop contribution to the up-type squark mass
(independently on the flavour direction) renders tachionic the squark eigenstate aligned with $\vec c$, and the spectrum is rejected.
Indeed, the points where there is a single light right-handed up squark with mass smaller than $1.5$ TeV, the red points, 
are at the border of the allowed region. 
The two loop contribution to the up squark mass in (\ref{def_thresold_R}) can be positive or negative, 
depending on the value of $\lambda$. This determines the shape of the allowed region, and
the location of the throat, which is approximately where the two loop correction in (\ref{def_thresold_R}) changes sign.

Note that in the red region only one mass eigenstate $\tilde u_1$ is light and much lighter than all the other squarks.
This mass eigenstate will be aligned along the vector $\vec c$ in flavor space, as the deformation $\lambda$ 
(if we neglect the 
effects of the CKM mixing).
This can be easily understood by observing that the gauge mediation contributions are diagonal in flavor space while the $\lambda$ deformation contributions are proportional to the 
matrix $c_{ij}=c_i c_j$, which has by construction $\vec c$ as eigenvector.
Hence the lightest right up-type squark will be a mixture of right handed stop, scharm and up-squark in the combination
\footnote{We used here the SLHA2 notations \cite{Allanach:2008qq} 
(i.e. $4 \leftrightarrow \tilde u_R \,, 5 \leftrightarrow \tilde c_R \,, 6 \leftrightarrow \tilde t_R \,$).}
\be
\label{mixing_conv}
\tilde u_1 = U_{14} \tilde u_R + U_{15} \tilde c_R+U_{16} \tilde t_R \qquad \text{with} \qquad |U_{14}|^2+ |U_{15}|^2+ |U_{16}|^2=1
\ee
where $|U_{14}| \simeq c_1,  |U_{15}|\simeq c_2,  |U_{16}|\simeq c_3$.

The contribution to the other squark mass $m_Q^2$ is also negative, 
but it is a two loop effect and it is compensated by the large value of the gauge mediation
contribution since we have considered large $\Lambda_S$.
This is also a welcome feature to alleviate possible sources of flavour violation in the $Q$ sector 
descending from the deformation $\lambda$, since the gauge mediation contributions are flavour diagonal.
The other generic effect of the deformation is that $m_{H_u}$ is very small at the scale $M$, and hence
it becomes very large and negative at the EW scale. As a consequence, since the EWSB condition sets approximately $\mu \simeq -m_{H_u}$,
the $\mu$ parameter is very large at the EW scale.
Hence the Higgsino is typically quite heavy.

In the plot on the right of Fig. \ref{lambda_scan} we focus on points with 
one light up-type squark with a mass smaller than $1.2$ TeV. We fix $\lambda=1$ and set $\Lambda/M$ in the range around $0.2$. Eventually we vary over $\vec c$. The plot shows the result of the scan on the 
$|U_{14}|, |U_{15}|$ 
mixing angle plane
(remember that $\sum_i U_{1i}^2=1$).

The red crosses are points which are not allowed because they do not respect flavour observables,
while the circular dots are viable points.
As mentioned above, the main flavour violating effect induced by the deformation $\lambda$ is in the RR up squark mass
matrix, precisely the one loop negative contribution in (\ref{def_thresold_R}). 
There is no off diagonal contribution to the $A$-terms, hence no LR mixing, and the two loop contributions to the LL squark masses induced by the 
deformation $\lambda$ are negligible compared to the flavour diagonal gauge mediated one (\ref{gmsb2}) for all the scanned points.
As a consequence the only relevant flavour constraint in Fig. \ref{lambda_scan} is actually coming from $D_0-\bar D_0$ mixing.
The plot shows that off diagonal contributions to RR up squark masses can be compatible with low energy flavour contraints 
if the mixing between the first and the second generation is not large. Hence in this model we can realize 
scenarios where the lightest squark is a highly mixed state in the flavor basis, either sup-stop or scharm-stop
(essentially the circular red points along the two axes).

In the plots, all points have a viable Higgs mass, which is a consequence of the fact that at least one of the two stops is very heavy, 
with a mass set mainly by $\Lambda_S$.
However the Higgs mass constraint implies that the deformation cannot lower too much the mass of the lightest eigenstate, 
if this has a significant component in the direction $3$ in flavor space. Hence very low mass values for the lightest squark are 
allowed only if they are not aligned along the third family.
This explains the hole of points in the region of small $c_1$ and $c_2$, where the deformation is mainly aligned along the direction $3$ in flavour
(note, we have fixed $\lambda=1$ and that $c_1^2+c_2^2+c_3^2=1$).
For a recent treatment of the Higgs mass formula in a fully flavored MSSM see e.g. \cite{Kowalska:2014opa}.

\begin{figure}[t]
\begin{center}
\includegraphics[width=0.5\textwidth]{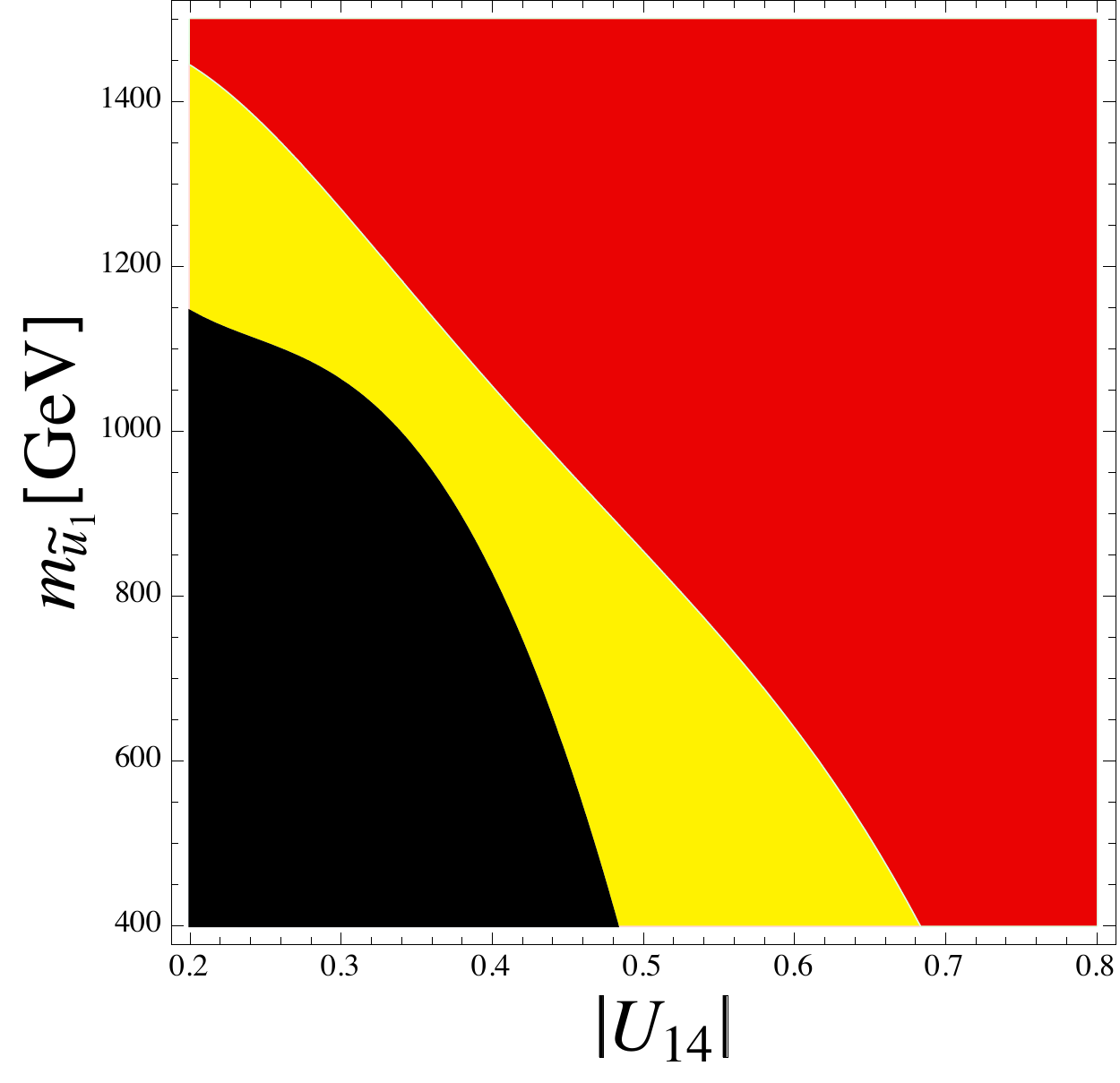}
\caption{\label{Higgs_mass}
Higgs mass constrains on the supersymmetric spectrum, assuming parameter as in Fig. \ref{lambda_scan}, (i.e. $\Lambda_S=1 \times 10^6$ GeV, $\Lambda_G=1.9 \times 10^5$ GeV, $\Lambda=5 \times 10^5 $ GeV), and where we furthermore set $\lambda=1$ and $c_2=0$. 
The red region has $m_h=125 \pm 2.5$ GeV, the yellow region $m_h = 120 \pm 2.5$ GeV, the black region even smaller $m_h$.
The mass of the lightest squark mass is denoted with
$m_{\tilde u_1}$ and $U_{14}$ is the sup-stop mixing angle (here $U_{15} \simeq 0$).
}
\end{center}
\end{figure}

This effect can be qualitatively estimated by considering Fig. \ref{Higgs_mass}, where we fix $c_2=0$ and we
plot the contours for the value of the Higgs mass (now allowed to take also small values) as a function of the lightest squark mass and of the mixing angle.
The region with viable Higgs mass is denoted in red.
The lighter the squark is, the more it should be aligned along the direction 1 in flavour space, and hence less in the direction 3 in flavour,
in order to have a large enough Higgs mass.

From Fig. \ref{Higgs_mass} it is evident that we can easily get viable points with a correct Higgs mass, satisfying flavor constraints, and having a
very light and highly mixed right-handed up squark.
In the scatter plots we fixed $\Lambda_G$ to a representative value, at the border of the LHC8 exclusion reach for the gluino mass;
however, the qualitative feature of the typical spectrum does not change by raising further the gaugino mass scale.
In Table \ref{benchmarks} we show two prototypical benchmark points, with different gluino mass, both presenting 
one lightest squark eigenstate, maximally mixed between stop and sup gauge eigenstates.

Summarizing, the deformation presented in this section provides a model realization of scenarios with a light squark with a large
stop-sup or stop-scharm mixing. The drawback of this model is that the Higgs mass is obtained by requiring the other squarks in the spectrum to be heavy,
implying a certain unavoidable amount of tuning in order to have one light squark state.

\begin{table}[t]
\begin{align}
&
\begin{tabular}{|c|c|c|c|}
\hline
$\Lambda_G$ & $M$ & $\lambda$ & $c_1$ 
 \\
 \hline
$1.9 \times 10^5$ GeV&  $ 2.31 \times 10^6$ GeV & 1.0 &0.75 
  \\
  \hline
 \end{tabular} \nonumber\\
 &
 \begin{tabular}{|c|c|c|c|}
 \hline
  $m_{\tilde u_1}$  & $|U_{14}|$& $|U_{15}|$ & $|U_{16}|$ \\
 \hline
 446 GeV &  0.706 & $\sim$ 0 & 0.707 \\
\hline
\end{tabular} \nonumber \\
&
\begin{tabular}{|c|c|c|c|c|c|}
\hline
 $m_{\tilde g}$ & $m_{\tilde W}$ & $m_{\tilde B}$ &  $m_h$ & $m_{3/2}$  \\
\hline
1.56 TeV & 526 GeV & 260 GeV & 124 GeV  &$1/k \times 612 $ eV  \\
\hline
\end{tabular}
\nonumber
\end{align}

\begin{align}
&
\begin{tabular}{|c|c|c|c|}
\hline
 $\Lambda_G$ & $M$ & $\lambda$ & $c_1$ 
 \\
 \hline
 $2.5 \times 10^5$ GeV&  $ 2.315 \times 10^6$ GeV & 1.0 &0.75 
  \\
  \hline
 \end{tabular}
 \nonumber
 \\
 &
 \begin{tabular}{|c|c|c|c|}
 \hline
  $m_{\tilde u_1}$  & $|U_{14}|$& $|U_{15}|$ & $|U_{16}|$ \\
 \hline
 657 GeV &  0.706 & $\sim$ 0 & 0.707 \\
\hline
\end{tabular} \nonumber \\
&
\begin{tabular}{|c|c|c|c|c|c|}
\hline
 $m_{\tilde g}$ & $m_{\tilde W}$ & $m_{\tilde B}$ &  $m_h$ & $m_{3/2}$  \\
\hline
1.99 TeV & 688 GeV & 345 GeV & 124 GeV  &$1/k \times 614 $ eV  \\
\hline
\end{tabular}
\nonumber
\end{align}

\caption{\label{benchmarks} Two example of viable spectra. 
The supersymmetry breaking scales for the scalar sector are fixed to $\Lambda_S=1 \times 10^6$ GeV and $\Lambda=5 \times 10^5 $ GeV.
The parameter $c_2$ is set to zero, hence $c_3^2=1-c_1^2$.
Concerning the rest of the spectrum, for both benchmarks: the sleptons are at least heavier than $1$ TeV; 
the other squarks and the higgsinos are further heavier, with masses larger than few TeV's.
}
\end{table}

\subsection{Mass Spectrum}
In this section we discuss the typical spectra that are generated by the model presented in the previous section.

We are interested in regions of the parameter space where there is a very light right-handed up squark, which is 
highly mixed in flavor.
In the following we focus on the case of large sup/stop mixing, assuming a small 
scharm component, consistently with the points shown in Fig. \ref{Higgs_mass}. We have shown in the previous
section that the complementary case of scharm/stop mixing can be also easily generated. The phenomenology of such case has been 
recently discussed in \cite{Blanke:2013uia}.

The supersymmetry breaking scale determining the gauge mediation contribution associated with the scalars ($\Lambda_S$)
is typically large in order to have the other squarks in the few TeV range. As a consequence, also the sleptons are very heavy and essentially 
decoupled from collider physics. The higgsinos are also heavy and decoupled, following the argument explained in section \ref{themodel}.

Hence the gauginos  and the gravitino are the only other supersymmetric particles that can play a role in the collider phenomenology
of these models.
In the model formulation, we made the assumption that we have only one gaugino mass scale $\Lambda_G$.
This implies that in the benchmark points shown in Table \ref{benchmarks} the gaugino masses 
respect the following relation
involving the gauge couplings
$
M_3 : M_2: M_1 = g_3^2 : g_2^2: g_1^2 
$.
However, we expect that this relation can be relaxed considering a complete GGM parameter space \cite{Meade:2008wd,Buican:2008ws}, 
effectively disentangling the
Bino from the Wino and from the gluino mass.
Note that modifying the gaugino mass at the messenger scale will affect the running of the sfermion masses, and hence could modify quantitatively
but not qualitatively the results we obtained above. 

The LSP is always the gravitino, whose mass is given by
$$
m_{3/2}=\frac{M \sqrt{\Lambda^2+\text{Max}(\Lambda_{G}^2,\Lambda_S^2)} }{\sqrt{3}  k M_{Pl} }
$$
and all the sparticles have universal decay to it with the formula
\be
\label{grav_decay}
\Gamma(\tilde X \to X \tilde G ) \simeq \frac{m_{\tilde X}^5}{48 \pi m_{3/2}^2 M_{Pl}^2}.
\ee
Here $k$ is a factor which is smaller than $1$, taking into account that the fact that the supersymmetry breaking
parameters coupling to the messengers can be smaller than the supersymmetry breaking scale of the complete model.
For instance, if the supersymmetry breaking scales $\Lambda,\Lambda_G,\Lambda_S$ are generated at one loop in some model of dynamical supersymmetry breaking,
the factor $k$ is a one loop suppression.
In the Table \ref{benchmarks} we show also the value of the gravitino mass for the prototypical benchmark point.
Depending on the value of $ k$, the decay of the NLSP can be displaced 
or longlived.
In the following we assume the second case, such that the Bino decay does not play any role in collider signals.


\section{Simplified model}\label{simplified_model}

\begin{figure}[t]
\begin{center}
\includegraphics[width=0.5\textwidth]{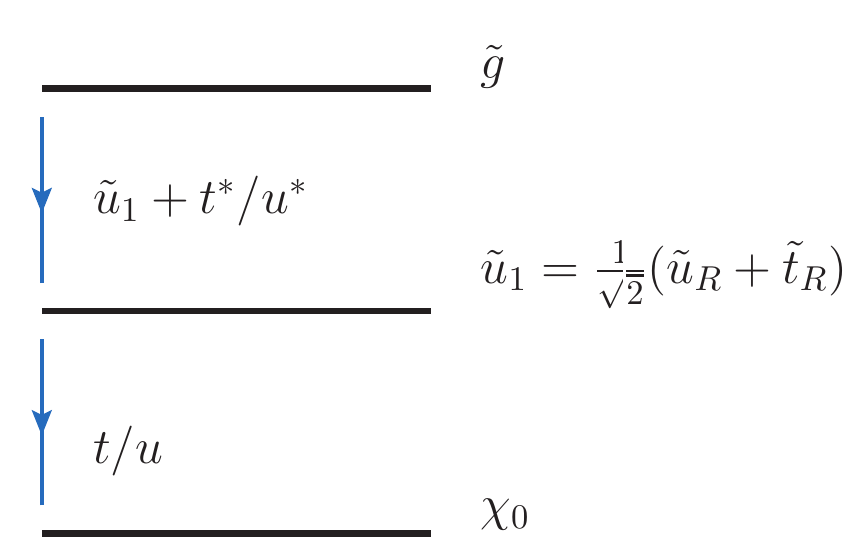}
\caption{\label{simplified} Maximally Mixed Sup-Stop (MMUT) scenario: the simplified model considered in the collider analysis.}
\end{center}
\end{figure}

Based on the previous analysis of the parameter space of the model, we here define the simplified spectra that we investigate
in the phenomenological study of the rest of the paper, and we set our benchmark points.
The only light supersymmetric states are a highly mixed right handed up-type squark, that we denote with $\tilde u_1$, with masses varying between 400 and 1000 GeV,
and a pure Bino neutralino $\chi_0$, with mass varying between 200 and $(m_{\tilde u_1}- 200)$. The gluino, which is relevant since it participates in the production mechanism of the light squark, and can also be directly produced, is fixed to $2$ TeV.
We focus on the case with maximal stop-sup mixing, i.e. the light squark state is a perfect mixture of sup and stop right-handed squarks  
$$
\tilde u_1=\frac{1}{\sqrt{2}} (\tilde u_R +\tilde t_R).
$$
The simplified model, that we dub as Maximally Mixed Sup-Stop (MMUT) scenario, is shown in Fig. \ref{simplified}, and is motivated by the spectrum structure that we described in the previous sections. 
In general, the Wino is considered to be heavy enough not to participate either in the squark production processes or in possible squark decays. 
\footnote{Actually EW production via Wino and also Bino exchange can increase the cross section at LHC of a few \% level, 
but we do not consider it in the following.}

\subsection{Branching Ratios}
We are interested in production of the lightest squark eigenstate and the resulting decay chain. 
Hence here we study the possible decay channel for $\tilde u_1$.

The decay of $\tilde u_1$ to the gravitino (given also by the universal formula (\ref{grav_decay})) 
is always suppressed compared to decays via gauge couplings.
If the mass difference between $\tilde u_1$ and $\chi_0$ is smaller than the top mass, $\tilde u_1$ will decay to
$u+\chi_0$ with a branching ratio which is almost $1$, as soon as there is a $u$ component inside $\tilde u_1$.
In this case its signature will resemble the one of a light up-type squark.
This can lead to interesting bounds in some regions of the squark-neutralino plane \cite{Agrawal:2013kha}.

A more unusual scenario can be realized when the mass difference between the $\tilde u_1$ and the $\chi_0$ is larger
than the top mass\footnote{Another interesting possibility is if $\chi_0$ is heavier than $\tilde u_1$, which can then decay only to the gravitino.}. We focus on this parameter region in the following sections. Here the two competing decays are $\tilde u_1 \to u \chi_0$ and $\tilde u_1 \to t \chi_0$.

\begin{figure}[t]
\begin{center}
\includegraphics[width=0.5\textwidth]{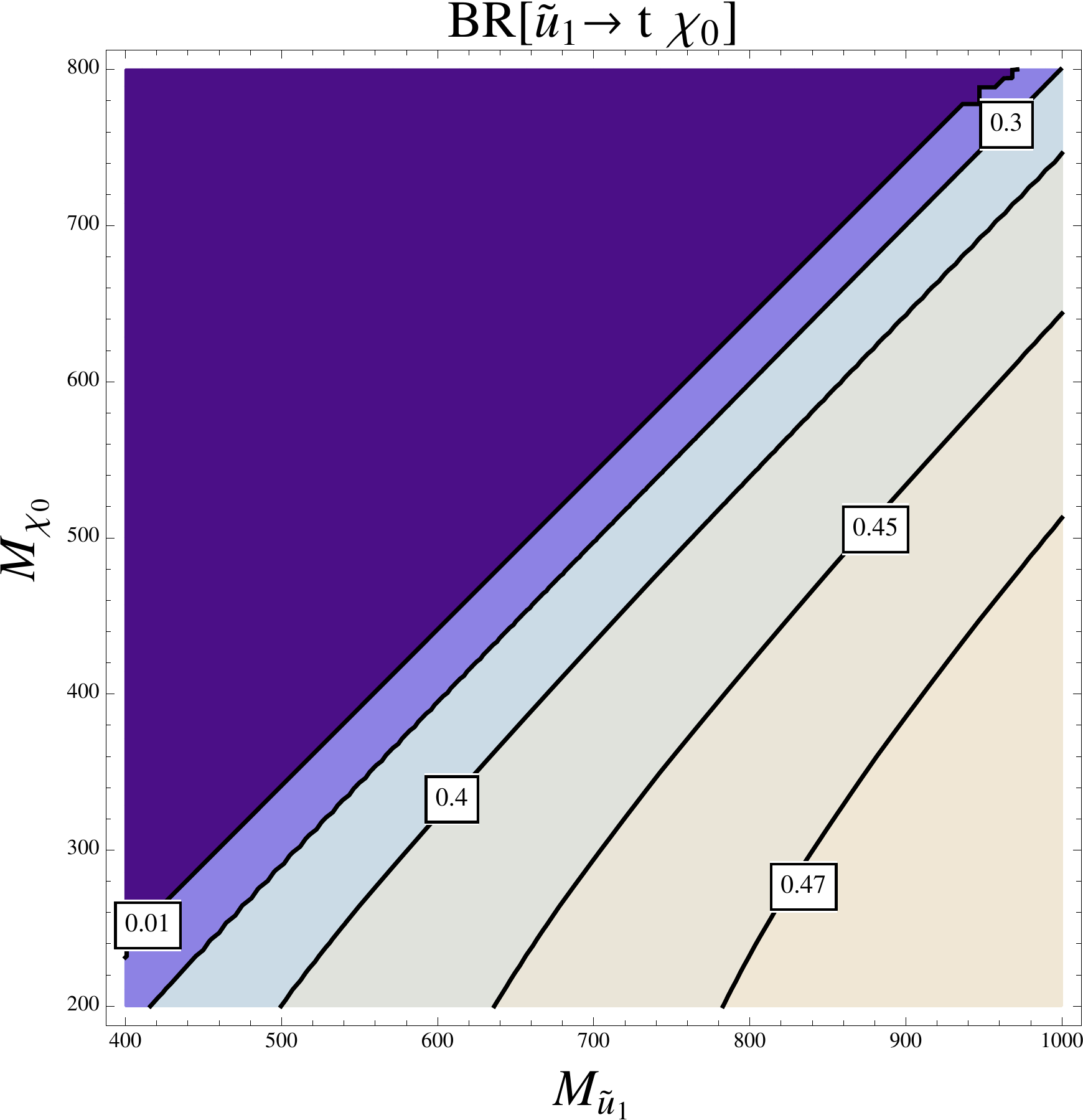} 
\caption{\label{BR} Branching ratio $BR(\tilde u_1 \to t \chi_0)$ as a function of the $\tilde u_1$ mass and of the neutralino mass.}
\end{center}
\end{figure}

The branching ratio for decays into top quarks is slightly suppressed even in the case of maximal mixing due to the phase space suppression from the large top mass. The formula for the branching ratio can be extracted by the analytic results of 
\cite{Bartl:1990ay,Boehm:1999tr}, and in the case of maximal mixing reads
\be
\label{eq:br}
BR[\tilde u_1 \to t \chi_0]=
\frac{\rho[m_{\tilde u_1}^2,m_{\chi_0}^2,m_t^2] \left(m_{\tilde u_1}^2-m_{\chi_0 }^2-m_t^2\right)}{\rho[m_{\tilde u_1}^2,m_{\chi_0}^2,m_t^2] ( m_{\tilde u_1}^2-m_{\chi_0 }^2-m_t^2) +\left(m_{\tilde u_1}^2-m_{\chi_0 }^2\right)^2},
\ee
where
\be
\rho[m_{\tilde u_1}^2,m_{\chi_0}^2,m_t^2]=\sqrt{m_{\tilde u_1}^4+ m_{\chi_0}^4+m_t^4-2 m_{\tilde u_1}^2 m_{\chi_0}^2-2 m_{\tilde u_1}^2 m_t^2 -2 m_{\chi_0}^2 m_t^2}.
\nonumber
\ee
In Fig. \ref{BR} we plot the branching ratio of Eq.(\ref{eq:br}) on the ($\tilde u_1$,$\chi_0$) mass plane.
Note that the branching ratio into top is still sizable even in region of moderate compression of the spectrum. 
For instance for $m_{\tilde u_1} =450$ GeV and $m_{\chi_0}=200$ GeV the
branching ratio into top is still between $30\%$ and $40 \%$. For an even more compressed region the branching ratio into tops drops abruptly.

Given the spectrum of the simplified MMUT scenario in Figure \ref{simplified}, at the LHC also the associated production of gluino and the lightest
squark state can play a significant role. The gluino can decay either as $\tilde g \to \tilde u_1 u^*$ or as $\tilde g \to \tilde u_1 t^*$ (and the conjugate processes),
hence giving extra jets or extra tops in the final state. Since the mass difference between the gluino (at $2$ TeV) and the lightest squark $\tilde u_1$ is 
large (at least larger than $1$ TeV on our parameter space), the phase space suppression induced by the top mass is always negligible. 
Indeed we checked that the branching ratio of the gluino decay into squark and up or top quarks is always proportional to the mixing angle.
Hence in the MMUT model of Figure \ref{simplified} the gluino will decay $50 \%$ to up-quark and $\tilde u_1$, and $50 \%$ to top-quark and $\tilde u_1$.

\subsection{Benchmark points}

We define five benchmark points on which we concentrate in the collider analysis, all with gluino mass fixed at $2$ TeV,
and with different squark and neutralino masses. They are reported in Table \ref{benchmark_points}.
The first one has very light squark but quite compressed spectra, with $m_{\tilde u_1} - m_{\chi_0} =250$ GeV. 
The others have larger squark masses and light or moderately heavy neutralino.
In the next sections we will discuss the exclusion limits on such benchmark at LHC8 and their distinctive signatures for LHC14, 
characterized by tops in the final state.

\begin{table}[h]
\begin{center}
\begin{tabular}{|c|c|c|c|c|}
\hline 
& $m_{\tilde g}$ & $m_{\tilde u_1}$ & $m_{\chi_0}$ & $\tilde u_1$ mixing angles \\
\hline
Benchmark Point 1 & $2$ TeV & $450$ GeV & $200$ GeV & $|U_{14}|=|U_{15}|=\frac{1}{\sqrt{2}}$\\
\hline
Benchmark Point 2 & $2$ TeV & $700$ GeV & $200$ GeV & $|U_{14}|=|U_{15}|=\frac{1}{\sqrt{2}}$\\
\hline
Benchmark Point 3 & $2$ TeV & $700$ GeV & $400$ GeV & $|U_{14}|=|U_{15}|=\frac{1}{\sqrt{2}}$\\
\hline
Benchmark Point 4 & $2$ TeV & $950$ GeV & $200$ GeV & $|U_{14}|=|U_{15}|=\frac{1}{\sqrt{2}}$\\
\hline
Benchmark Point 5 & $2$ TeV & $950$ GeV & $400$ GeV & $|U_{14}|=|U_{15}|=\frac{1}{\sqrt{2}}$\\
\hline
 \end{tabular}
 \caption{\label{benchmark_points} Benchmark points considered in the collider study of the simplified MMUT model.}
 \end{center}
 \end{table}

\section{Collider signatures}\label{LHC_section1}

\subsection{Production modes and cross sections}
\begin{figure}[t]
\begin{center}
\includegraphics[width=0.45 \textwidth]{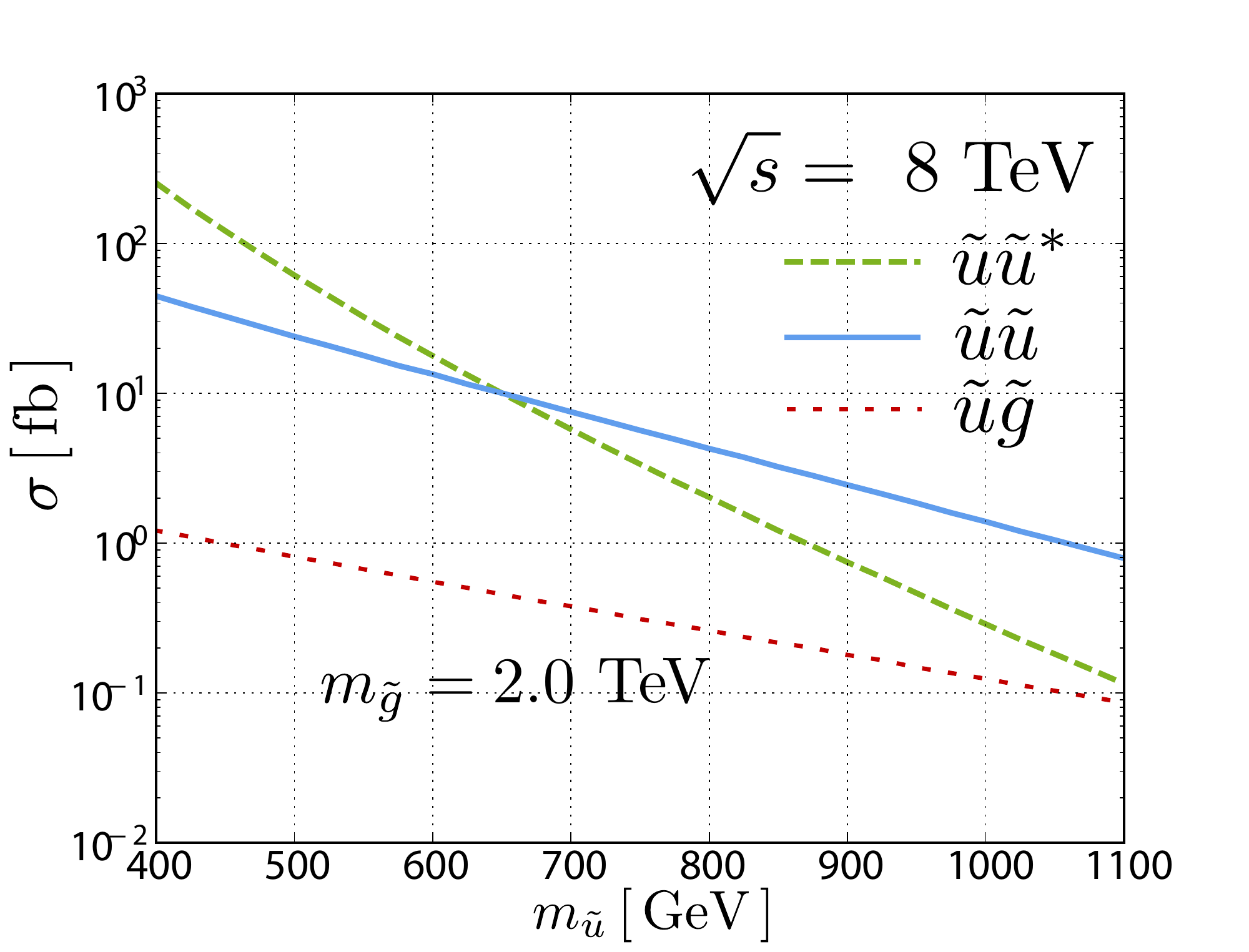}
\includegraphics[width=0.45 \textwidth]{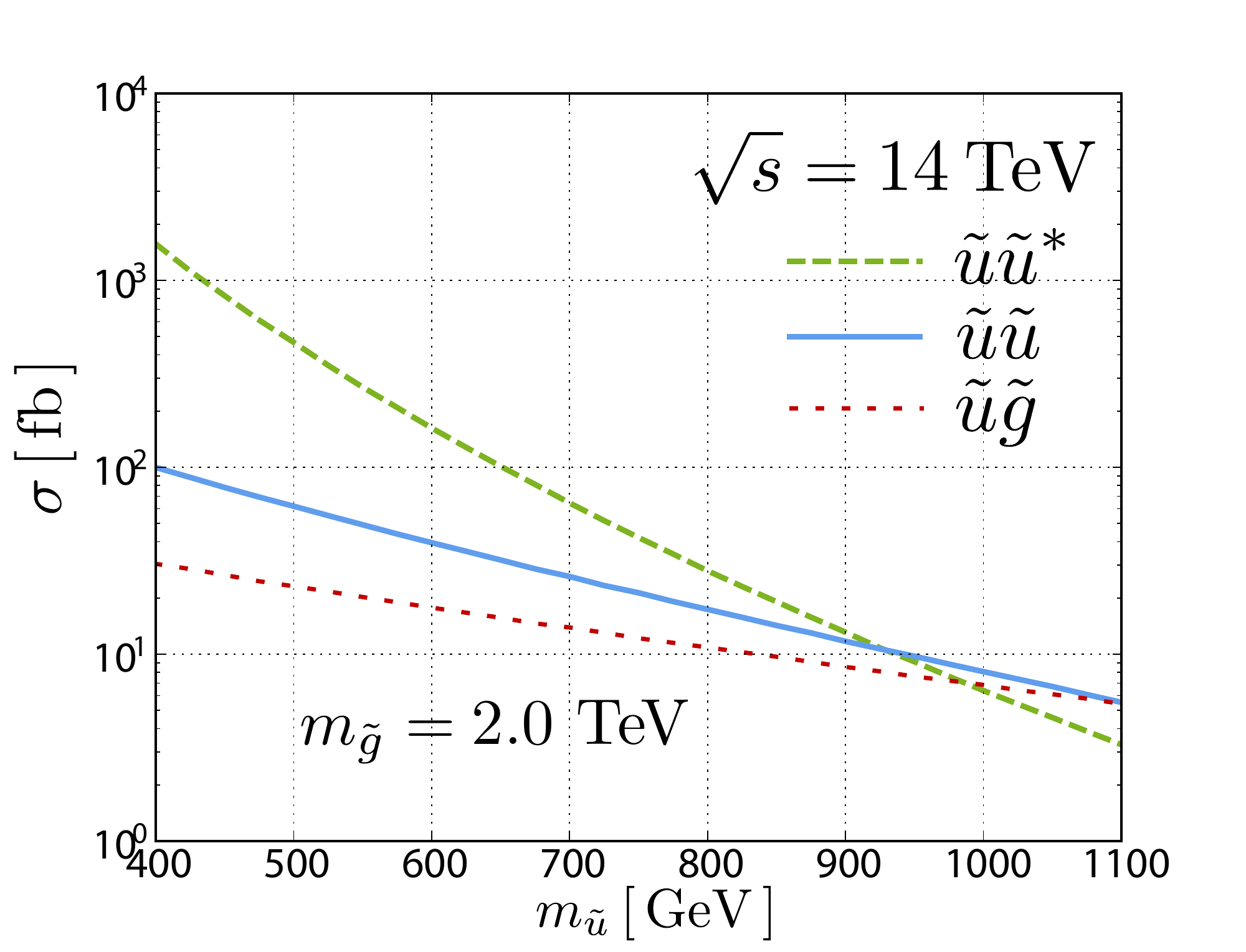}
\caption{\label{xsec}  Cross sections in the MMUT model for squark-antisquark,  squark-squark and gluino-squark production at LHC8 and LHC14
as a function of the squark mass. The gluino mass is fixed to $2$ TeV.}
\end{center}
\end{figure}

Given the simplified spectrum of the MMUT scenario in Fig. \ref{simplified}, at the LHC we expect the following production modes:
\be
p p \to \tilde u_1 \tilde u_1^* \qquad , \qquad p p \to \tilde u_1 \tilde u_1 \qquad , \qquad p p \to \tilde u_1 \tilde g
\ee
Cross sections for squark-antisquark, squark-squark and gluino-squark production at LHC8 and LHC14 are shown in Fig. \ref{xsec},
computed at LO using \verb|MadGraph 5|~\cite{Alwall:2011uj}.

Note that gluino pair production, gluino-antisquark production and antisquark pair production
do not contribute significantly to the SUSY production modes, because of the PDF suppressions.
To give an estimate of these contribution and compare with the cross sections depicted in Figure \ref{xsec} 
we evaluate these cross section with $m_{\tilde u} = 400$ GeV and $m_{\tilde g}=2$ TeV.
The cross section for gluino pair production is $\sigma \sim 10^{-3}~\text{fb}$ and $\sigma \sim 0.7~\text{fb} $ at 8 TeV and 14 TeV respectively, while the cross section for gluino anti-squark production is $\sigma \sim 0.02~\text{fb}$ and $\sigma \sim 1~\text{fb}$.
For the anti-squark pair production cross section we find $\sigma \sim 0.2~\text{fb}$ and $\sigma \sim 1.6~\text{fb}$ at 8 TeV and 14 TeV respectively.
Hence they are negligible and we will not consider them in our analysis.

We stress that the production modes of squark-squark and gluino-squark are sizeable only because of the up component in the lightest squark,
and thus they are weighted by the mixing angle. For instance the squark-squark production is proportional to $|U_{14}|^4$, which is $1/4$ in our maximally mixed scenario.
This production mode vanishes in the scenario where the lightest squark is purely stop-like, and is maximized when the lightest squark is a pure up squark.

\begin{figure}[t]
\begin{center}
\includegraphics[width=0.48\textwidth]{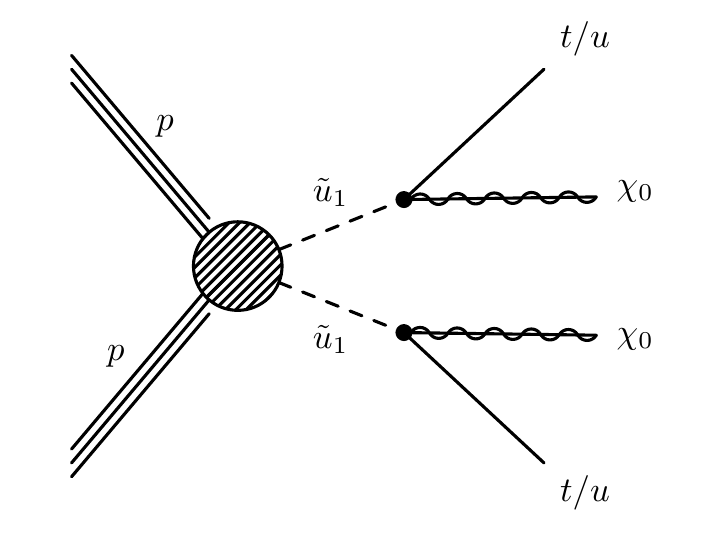}
~~
\includegraphics[width=0.48\textwidth]{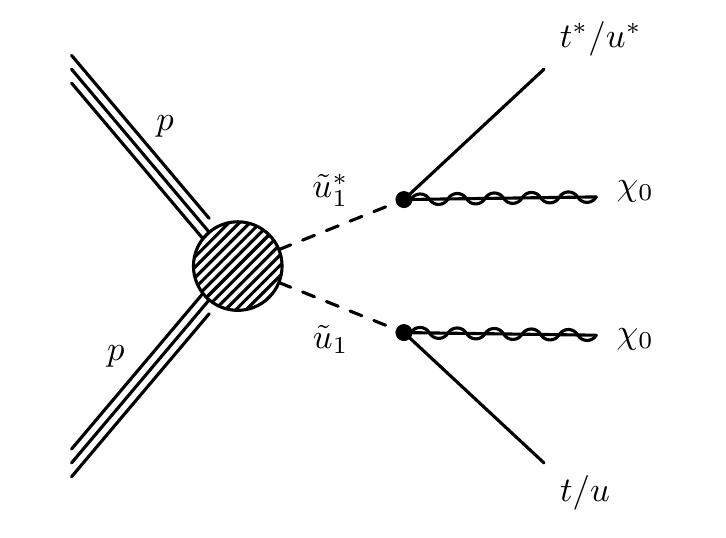}
\includegraphics[width=0.48\textwidth]{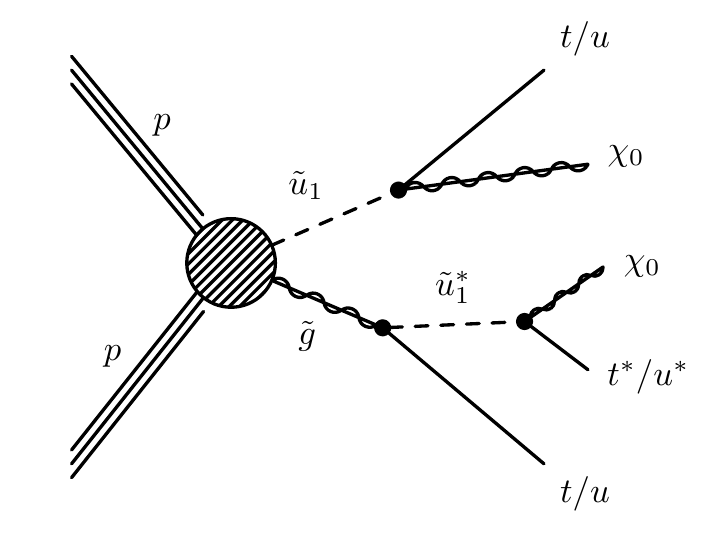}
\caption{\label{sup1_pairs} Top left: squark pair production
(note that this process is produced through t-channel exchange of gluino). 
Top right: squark antisquark pair production. 
Bottom: squark gluino production (also the conjugate decay process for the gluino decay should be included).
}
\end{center}
\end{figure}

Including the possible decay modes, the relevant production mechanisms are depicted in Figure \ref{sup1_pairs}.
Depending on the decay of the gluino and of the lightest squark, the final state can be composed of 
up quarks or of top quarks and neutralinos. Since we consider the neutralino to be stable in terms of collider time scales, it will give rise to missing energy signatures.

One can generate the following signals from squark-antisquark pair production (right Fig. \ref{sup1_pairs})
 \be
 \label{squark_antisquark}
 pp \to \tilde u_1 \tilde u_1^* \to ( j j \chi_0 \chi_0,  j t \chi_0 \chi_0,  t t^* \chi_0 \chi_0 )
 \ee
the following final states from squark-squark pair production (left Fig. \ref{sup1_pairs})
\be
\label{squark_squark}
p p \to \tilde u_1 \tilde u_1 \to ( j j \chi_0 \chi_0 , , j t \chi_0 \chi_0 , t t \chi_0 \chi_0)
\ee
and the following signatures from squark-gluino associated production (bottom Fig. \ref{sup1_pairs})
\be
\label{gluino_squark}
p p \to \tilde u_1 \tilde g \to (j j j \chi_0 \chi_0, j j t \chi_0 \chi_0, j j t^* \chi_0 \chi_0, j t t^* \chi_0 \chi_0, j t t \chi_0 \chi_0, t t t^* \chi_0 \chi_0 )
\ee
Besides the usual supersymmetric signatures of jets plus $\MET$ or of top-antitop pair plus $\MET$,
among the possible final state we can
find the single top (hence single lepton) and same sign tops (hence two positive same sign leptons).
Especially this second signature, i.e. same sign tops, is a unique consequence of the maximal flavour mixing in our MMUT scenario,
and can be considered as a probe of the mixing angle of the lightest squark state.

\subsection{Constraints from LHC8}
The supersymmetric processes described above contribute to final states with jets and missing transverse energy, and are possibly probed by 
the jets plus $\MET$ searches of ATLAS \cite{TheATLAScollaboration:2013fha} and CMS \cite{CMS:2014ksa} at 8 TeV.
Analogously, presenting top pair and $\MET$ in the final states, they could be also probed by the standard stop
searches of ATLAS \cite{Aad:2014kra} and CMS \cite{Chatrchyan:2013xna}.

We thus have to verify that our benchmark points in Table \ref{benchmark_points} are not excluded by existing LHC8 searches. In order to ascertain the viability of
our benchmark points we produce samples with \verb|MadGraph 5|, shower them with \verb|Herwig++| \cite{Bahr:2008pv}, 
and process them using \verb|CheckMate| \cite{Drees:2013wra},
using all the available LHC searches, including jets plus $\MET$ \cite{TheATLAScollaboration:2013fha}, searches exploting the $\alpha_T$ variables \cite{Chatrchyan:2013lya},
and single-lepton stop searches \cite{Aad:2014kra}.
We find that all five benchmark points in Table \ref{benchmark_points} are still not excluded by LHC8 searches.
The benchmark points which are at the border of the exclusion are the first and the second one in Table \ref{benchmark_points}.
However, with an overall $K$ factor as large as\footnote{Considering the relevant production modes involved in our process, a K-factor of $1.7$ is 
the largest possible value at 8 TeV \cite{Goncalves:2014axa,GoncalvesNetto:2012yt}.} 
$1.7$, the CheckMate analysis concludes that these two benchmark points are still allowed.
The stronger constraints come from the ATLAS jets plus $\MET$ search \cite{TheATLAScollaboration:2013fha}.

This result could be surprising given that the SUSY cross section is enlarged by the contribution of the squark-squark production mode
with respect to the unmixed squark scenario, where the only relevant contribution is coming from squark-antisquark production (the gluino-squark 
associated production is negligible at $8$ TeV in the small squark mass region).
However, there are two effects which both contribute in making the signal of the MMUT model difficult to exclude.
First, the cross section enhancement due to the squark-squark channel is only mild, given that this production mode is 
suppressed by the mixing angle to the fourth power; this results in a factor of $1/4$ suppression, as mentioned above (see Figure \ref{xsec}).
Second, the different decay modes of the light squark eigenstate, including top quarks, make the signal less clean than pure jets plus $\MET$ topologies.
This is particular effective in the first benchmark point (with $m_{\tilde u_1}=450$ GeV and $m_{\chi_0}=200$ GeV). 
In this case the branching ratio of the lightest squark into top is still sizeable (see Figure \ref{BR}),
but the spectrum is moderately compressed and the resulting tops will be quite soft, effectively reducing the efficiencies.

\begin{figure}[t]
\begin{center}
\includegraphics[width=0.8 \textwidth]{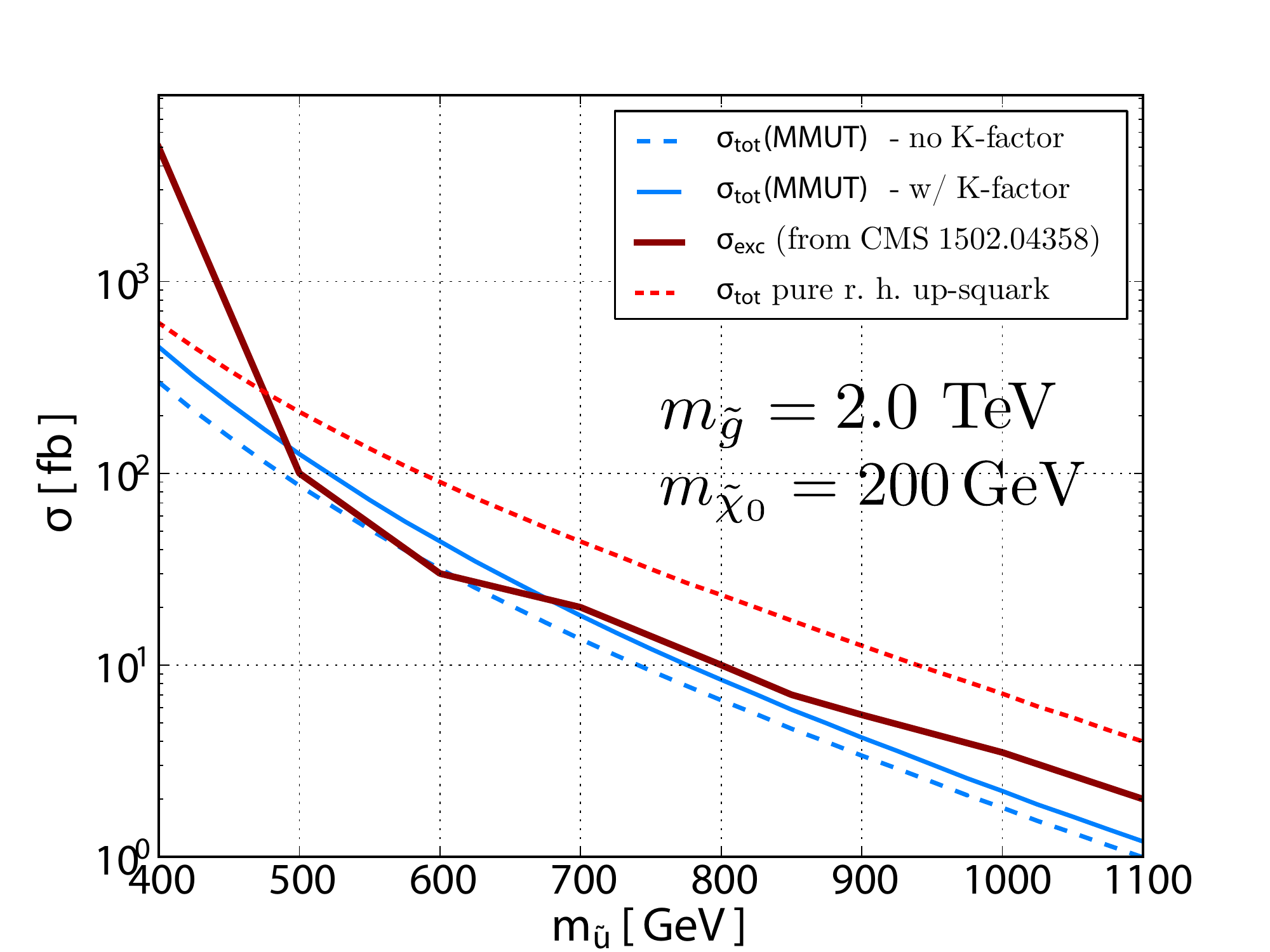}
\caption{\label{exclusion_plot} Simplified exclusion plots based only on the SUSY production cross section. The neutralino mass is fixed to $m_{\chi_0}=200$ GeV and the gluino 
mass to $2$ TeV. The dashed, blue curve is the LO total SUSY production cross section for the MMUT model, including all relevant production channels ($\tilde u_1 \tilde u_1^*,\tilde u_1 \tilde u_1,\tilde u_1 \tilde g$). The solid, blue curve is the same total cross section for the MMUT model, but re-weighted by the appropriate $K$-factors; we used $K_{\tilde u_1 \tilde u_1^*}=1.6$,  
$K_{\tilde u_1 \tilde u_1}=1.1$ and $K_{\tilde u_1 \tilde g}=1.7$, taken from \cite{Goncalves:2014axa,GoncalvesNetto:2012yt}. 
The solid, dark red curve is the exclusion cross section extracted from the
CMS analysis \cite{Chatrchyan:2013xna} for neutralino mass at 200 GeV.
The dotted, red curve, for comparison purposes, is the cross section for the case of a pure (right handed) up squark, also re-weigthed with the appropriate $K$-factors. 
}
\end{center}
\end{figure}

In order to give an intuitive explanation of these features, in Figure \ref{exclusion_plot} we plot the total cross section of our supersymmetric production modes,
weighted with K-factors taken from \cite{Goncalves:2014axa,GoncalvesNetto:2012yt}, and the cross section exclusion limit extracted from CMS \cite{Chatrchyan:2013xna} 
in the case of neutralino mass at 200 GeV.
This plot is an oversimplification, since it does not take into account the different decay modes and efficiencies in the various channels.
Nevertheless it allows to get some insight on the LHC8 reach for the MMUT simplified model and how the mixing angle 
plays a crucial role in reducing the experimental bounds.
We observe that the region of squark masses between 500 GeV and 680 GeV is excluded, with the 450 GeV and the 700 GeV benchmarks 
just above the experimental reach, consistent with our results from CheckMate.
To provide a further quantitative estimate of the effectiveness of the mixing angle suppression, 
in the plot we also show the total production cross section for a right handed up squark, 
not mixed. In this case the same sign production mode is not suppressed by the mixing angle and the total cross section is considerably larger,
leading potentially to very strong bounds. In our plot, the exclusion limit for scenarios with a  pure up-type right handed squark seems to reach very large values for the
squark mass; this is a consequence of the oversimplification of this estimate, for a complete recasting and evaluation of the LHC reach see \cite{Mahbubani:2012qq}.

Other searches that could potentially constraint the model are the standard supersymmetry searches for stops \cite{Aad:2014kra,Chatrchyan:2013xna}.
However, note that the total production cross section in our scenario is smaller than in the case of a pure light stop.
Indeed, even if in addition to the squark-antisquark production mode we have also the squark-squark production mode, the entire cross section
should be multiplied by a factor of $BR(\tilde u_1 \to t \chi_0)$, in order to require at least one top in the final state. 
One can easily check that for instance on the first benchmark point (with $m_{\tilde u_1}=450$ GeV and $m_{\chi_0}=200$ GeV)
this implies that the cross section times the branching ratio for a single top in the MMUT model would
be eventually lower than the single top production in a model with a light pure (right handed) stop with the same mass.
Hence, these simplified estimates already provide an understanding which is consistent with the robust result obtained through Checkmate.
For squark masses larger than 700 GeV we do not expect existing stop searches to be relevant.


For squark masses smaller than 450 GeV, the branching ratio into tops drops significantly, to get completely negligible already at a squark mass of 420 GeV (see Figure \ref{BR}).
Moreover, the spectrum is considerably compressed and hence weakly constrained by LHC searches.
Hence in the MMUT model there is a small window for squark mass around 450 GeV which is still allowed by LHC8 searches and which present tops in the final state,
thanks to the peculiar properties associated with maximal flavour mixing. In the following, we find therefore interesting to include this blind-spot benchmark point in our analysis and discuss its phenomenology at LHC14.

The above discussion is valid for a neutralino mass of $200$ GeV. For lower neutralino masses, the bound on the squark mass is stronger. Conversely, a heavier neutralino mass leads to weaker bounds.

To summarize, the five benchmark points for the MMUT model we defined in Table \ref{benchmark_points} are still allowed by actual LHC8 searches.
Nevertheless, we expect that in the next LHC run at higher energy the standard SUSY searches for light squarks or stops could eventually probe them.
In the following we concentrate on the new signatures associated with large flavour mixing, and access the LHC14 reach for such signatures.
These are new interesting channels to look for supersymmetric scenarios, and at the same time represent a powerful strategy to test the mixing property of the light squark.

The collider signatures distinctive of our simplified models are:

\begin{enumerate}
\item Single-top signal arising from all production modes (\ref{squark_antisquark},\ref{squark_squark},\ref{gluino_squark})
\item The new signature of the sup-stop mixing, i.e. same sign-tops, arising from the production modes (\ref{squark_squark}) and (\ref{gluino_squark})
\end{enumerate}
In the next section we will study the prospects for these two new channels at LHC14 for the five benchmark points of the MMUT simplified model.

\section{Top signatures at LHC14}\label{LHC_section2}

\subsection{Event Generation and Reconstruction}

We generate all events using leading order \verb|MadGraph 5|~\cite{Alwall:2011uj} with the \verb|NNPDF 2.3|~\cite{Ball:2012cx} set for the parton distribution functions. Upon hard level process generation, we further shower the events using \verb|Pythia 6|~\cite{Sjostrand:2006za}. We match the background event samples, where relevant, to extra jets using the MLM matching scheme~\cite{Mangano:2006rw}, with \verb|QCUT=30 GeV| and \verb|xqcut = 20 GeV| in case of top production and \verb|QCUT=15 GeV| and \verb|xqcut = 10 GeV| for production of weak bosons. 
Our analysis includes detector effects on event reconstruction, where we utilise \verb|Delphes 3|~\cite{deFavereau:2013fsa} with the default CMS settings for event reconstruction, $b$-tagging efficiencies and lepton isolation. 

In the following sections we consider signatures of our model in the single top (i.e. $l + \MET + b$) channel as well as the same sign positive top (i.e. $l^+ l^+ + \MET$) channel, requiring us to consider a range of SM background channels. For the purpose of studying the collider reach in the single top channel, we consider $t\bar{t}$ and $W+b\bar{b}$ events matched to one extra jet as well as $tW$ events, where we require at least one lepton of unspecified charge at hard process generation level. 

Our SM background samples for the same sign lepton analysis consist of $t\bar{t}$ and $W+b\bar{b}$ events matched to one extra jet, where we require one positively charged lepton at generator level. In addition, we also consider rare SM processes where we include production of  $t\bar{t}W, t\bar{t}Z, ZZ, W^+ W^+ W^-$ and $WZ$. Here we require at least one positive lepton in the final state.

In order to improve our estimates of background event yields, we normalise the $t\bar{t}$ production cross sections to the NNLO+NNLL value of Ref.~\cite{Czakon:2013goa}, while we assume a conservative K-factor of 1.4 for $W+b\bar{b}$ and single top production and $1.3$ for rare SM processes. For the purpose of signal generation, we always assume a K-factor of 1.1 for $\tilde{u}\tilde{u}$, 1.4 for $\tilde{u}\tilde{g}$ and 1.5 for $\tilde{u}\tilde{u}^*$ production~\cite{Goncalves:2014axa,GoncalvesNetto:2012yt}.

\subsection{Single-top Channel}
\input{single_top.tex}

\subsection{Same sign top Channel}

\input{same_sign.tex}

\section{Conclusions}
The canonical paradigm of supersymmetry typically assumes ultra-violet completions with 
a flavour symmetry, thereby leading to aligned or diagonal soft masses.
Yet, the possibility of large off diagonal entries represents a viable and interesting option 
to be investigated in the context of the MSSM.

In this paper we studied an example model of supersymmetry with large mixing in the right-handed up-type squark mass matrix in the context of extended  gauge mediation. Our analysis includes detailed consideration of  constraints imposed on the parameter space
by flavour observables, as well as prospects for LHC Run II to discover models with large squark mixings. 
%
We find that the single-top final signatures predicted by our model can be accessible at LHC14 with 300 fb$^{-1}$. A more distinctive feature of the MMUT model, the same sign positive lepton final state which signals a large squark mixing, can be probed at the high luminosity LHC. A combination of the two searches could provide useful insight into the flavor mixing properties of the light squark state.

As models with large squark mixings typically suffer from an increased degree of fine tuning, we take a moment to discuss the naturalness of the MMUT model.
The parameters determining the tuning in the MSSM are the dimensionful terms entering in the corrections of the up type higgs soft mass:
\be
16 \pi^2 \frac{d }{dt} m_{H_u}^2 = 6~ \text{Tr} \left( y_u^{\dagger} m_{Q} y_u + y_u^{\dagger} m_U y_u +A_u^{\dagger} A_u \right) +\dots  \non \,,
\ee
eventually leading to a
fine tuning in the EWSB correction. 
From this expression, one concludes that the relevant entries of the squark mass matrix in the evaluation of the tuning
are the ones projected onto the Yukawa directions, essentially the $(3,3)$ entries.

The reference scenario to which we compare the degree of fine tuning is natural SUSY \cite{Papucci:2011wy}, 
where the lightest right handed squark 
is a pure stop, not mixed in flavor. 
The tuning in natural SUSY is set by the LHC bound on right handed stop which, in the case of 200 GeV neutralino, is around $m_0 \approx 650 \GeV$ \cite{Aad:2014kra}. 
In analogy with \cite{Blanke:2013uia}, we define a parameter measuring the departure from natural SUSY by 
dividing the $(3,3)$ entry in our MMUT scenario with the minimal one of natural SUSY:
$$
\xi \equiv \frac{m_{U_{3,3}}^2}{m_0^2}=  \frac{ |U_{1,6}|^2 m_{1}^2 + |U_{2,6}|^2 m_2^2}{m_0^2} = \frac{ \frac{1}{2}m_{1}^2 + \frac{1}{2} m_2^2}{m_0^2}\,,
$$
where we labelled with $m_1$ and $m_2$ the lightest and next to lightest up-squark eigenstate, respectively.

In the simplified model of Section \ref{simplified_model} we considered only one of the lightest eigenstates. 
We found that the lower bound on such state from LHC 8 TeV searches is around $700$ GeV
(neglecting here the possibility of 
$m_{\tilde u_1}=450$ GeV which is a peculiar very compressed point; see Fig. \ref{exclusion_plot}).
In order to provide a quantitative estimate of $\xi$, we should specify also the value of the other up-squark eigenstate.
In Section \ref{simplified_model} we assumed that the next to lightest squark eigenstate is decoupled from
LHC physics, and taking inspiration from the gauge mediation model of Section \ref{themodel}, we can consider it to be 
at the $O(5-10)$ TeV scale. In this case the MMUT model would be considerably more tuned than natural SUSY by a factor $\xi \simeq 30 - 100 $.
In the most optimal scenario, instead, we can assume that the same LHC bound  on the lightest squark eigenstate
applies also to the next to lightest eigenstate. In this case
\footnote{Note that here one should rely on some extra mechanism (such as the NMSSM) to obtain the correct Higgs mass.}
we apply a common bound of $m_{\tilde u_1} \sim m_{\tilde u_2} \geq 700$ GeV,
and the tuning of the MMUT model with respect to natural SUSY reduces to $\xi \simeq 1.2$.
Hence we find that in the optimal case, the MMUT model also represents a slightly more un-natural SUSY scenario.  The reason is that the LHC bound on the lightest squark state in the MMUT model
 is higher than the LHC bound on a pure stop eigenstate, since in the former case we have extra production modes (in particular process 1 in Fig. \ref{sup1_pairs}; see
 also the cross Section plot of Fig. \ref{exclusion_plot}).
We  conclude that generically the MMUT model will be at least slightly more tuned than natural SUSY.

Future studies of models with large squark mixings  would benefit from including  the possible scenario where the neutralino
is not stable on collider scales, adding typically two extra displaced photons to the signatures we discussed, which could potentially be accessed at the LHC Run II.


On the model building side, we note that we did not address the issue about the dynamical origin of supersymmetry breaking.
It would be interesting to explore this aspect at greater depth, as well as to evaluate the model's effective level of tuning from a UV perspective.
%
%


\section*{Acknowledgments}
We are grateful to Gilad Perez for discussions and insightful comments on a preliminary version of this paper. 
We also thank David Shih and Jared Evans for useful comments on the draft.
A.M. would like to thank also Riccardo Argurio, Lorenzo Calibbi, Athanasios Dedes, Didar Dobur and Diego Redigolo for interesting discussions.
 M.B. would like to thank the University of Kansas phenomenology group for their hospitality during the final stages of the project.
M.B. and A.M. are supported in part by the Belgian Federal Science Policy Office through the Interuniversity Attraction Pole P7/37. 
A. M. is a Pegasus FWO postdoctoral Fellowship. A. M. is also supported in part by the Strategic Research Program High Energy Physics
and the Research Council of the Vrije Universiteit Brussel.

\appendix

\section{Soft terms from R-symmetric hidden sector}\label{appendixA}

In this section we obtain the soft term contribution for the model studied in this paper, i.e.
an hidden sector with a discrete $\mathbf{Z}_4$ R-symmetry (\ref{R-preserving}) coupled via superpotential interaction (\ref{deformation}) to the up type quark.

The cases of non-R symmetric hidden sectors coupled via messenger matter coupling to the MSSM have been completely classified in \cite{Evans:2013kxa}.
The coupling with the same structure we are considering has been denoted as $I13$ in their classification, and it consists of
the messenger-matter coupling
\be
W= \lambda \sum_{i=1}^3 c_i U_i \phi_1 \phi_2
\ee 
where the messenger are assumed to be part of a supersymmetry breaking sector coupling to a spurion as
\be
W_{SUSYbr}= X (\tilde \phi_1 \phi_1 +\tilde \phi_2 \phi_2) 
\ee
with $X=M+\theta^2 F$.

In this case the contribution to the soft masses can be extracted from the formulas of \cite{Evans:2013kxa} and results
{\small{
\bea
\label{def_thresold_nonR}
&&
A_{U_i F_{U_j}}= -\frac{d_U}{16 \pi^2} \lambda^2 c_{ij} \Lambda  \\
&&
\delta m_{U_{i} U_{j}}^2= \frac{1}{256 \pi^4} c_{ij} \lambda^2 d_U  \left(  \lambda^2 d_{\phi}  +   \lambda^2 d_U  -2    \sum_{r=1,3} C_r g_r^2 \right) \Lambda^2
-\frac{d_U}{48 \pi^2} c_{ij} \lambda^2  h(\frac{\Lambda}{M}) \frac{\Lambda^4}{M^2} \nonumber \\
&&
\delta m_{Q_{ij}}^2=-\frac{d_U \lambda^2}{256 \pi^4} (y^{\dagger}.c.y)_{ij} ~\Lambda^2 \nonumber  \\
&&
\delta m_{H_u}^2=-\frac{3 d_U \lambda^2}{256 \pi^4}  \Tr (y^{\dagger}.c.y) ~\Lambda^2 \nonumber 
\eea}}
\noindent
where $d_U=2$ and $d_{\phi}=4$, $C_1=2/5,C_2=0,C_3=4$, $c_{ij}= c_i c_j$ 
and we used that $ c_1^2+c_2^2+ c_3^2=1$.

These results have been obtained by studying the threshold corrections to the wave function renormalization of the MSSM matter fields
induced by integrating out the messengers. Precisely, in the case of the up type quark (the other sfermions are analogous), 
the results (\ref{def_thresold_nonR}) arise from the correction in the Kahler
potential
\be
\int d^4 \theta Z_{u_{i} u_{j}} (|X|) U_{i}^{\dagger} U_{j} \supset  |F|^2 (\partial_X \partial_{X^*} \Zuu) U^{\dagger}_i U_j +\big[ F (\partial_X  \Zuu) F_{u_i}^{\dagger} U_j+h.c. \big ]
\ee
leading to
\bea
&&
A_{U_{i}U_{j}}=F (\partial_X  \Zuu) \\
&&
m_{U_i U_j}^2=-|F|^2 (\partial_X \partial_{X^*} \Zuu) + |F|^2 (\partial_X  Z_{u_{i} u_{k}})  (\partial_{X^*}  Z_{u_{k} u_{j}}) 
\eea
where the second term in the mass squared arises from integrating out the F term of the MSSM field.
From this procedure it is clear that the contributions to the soft masses (at two loop) can be divided into a contribution coming 
from the second derivative of $Z_{u_{i} u_{j}}$ with respect to $X$ and $X^*$,
and another one coming from the square of the first derivative of $\Zuu$, i.e. the A-term squared. 
Indeed the authors of \cite{Evans:2013kxa} splitted the contributions to $m_{soft}^2$ schematically into $m_{soft}^2=\hat m_{soft}^2 +|A|^2$ to make
it explicit. In the formulas (\ref{def_thresold_nonR}) the A-term contribution to the soft masses is the second term in the big round parenthesis. 

The last term in $\delta m_{U_{i} U_{j}}^2$ in (\ref{def_thresold_nonR}) is coming from one-loop correction to the soft mass, it is not at leading order in $F/M$ but plays a crucial role in
lowering the squark mass. The precise expression for the $h$ function is
\be
h(x)=\frac{3}{x^4} \Big(  (x-2) \ln(1-x) -(x+2) \ln (1+x) \Big)
\ee
and it is an even function of $x$.

Now, in order to obtain the complete set of soft terms induced in our model by the R-symmetric SUSY breaking sector (see eqn. (\ref{R-preserving})), 
we adopt the following strategy. Consider a double copy of the above non R-symmetric model, with two different spurions $X_1$ and $X_2$, but with the same coupling
to the MSSM matter field
\be
\label{complete1}
W=\lambda \sum_{i=1}^3 c_i U_i (\phi_1 \phi_2 +\phi_3 \phi_4)+X_1 (\phi_1 \tilde \phi_1 +\phi_2 \tilde \phi_2) + X_2   (\phi_3 \tilde \phi_3 +\phi_4 \tilde \phi_4) 
\ee
As in the case above, the one loop corrections to the A-terms and the two loop corrections to the soft masses are encoded into the wave function
renormalization for the MSSM field which receive in this case two additive contributions
\footnote{The two sectors are coupled only via the up type quarks and the correction of one sector to the other are loop suppressed and negligible, 
unless there is a huge hierarchy between $X_1$ and $X_2$. 
This is analogous to the situation for two SUSY breaking hidden sector in gauge mediation \cite{Argurio:2011hs}.}
\be
\Zuu = \Zuu(|X_1|)+ \Zuu(|X_2|)
\ee
Note that the functional form of $\Zuu (|X_1|)$ and $\Zuu(|X_2|)$ is the same.  

The correction induced by the Kahler potential are then
\bea
\int d^4 \theta Z_{u_{i} u_{j}} U_{i}^{\dagger} U_{j} \supset && \Big[ 
\Big(  F_{1} \partial_{X_1} \Zuu(|X_1|) +F_2 \partial_{X_2} \Zuu(|X_2|) \Big) F_{u_i}^{\dagger} U_j + h.c. \Big] 
\nonumber \\
&&
+\Big(  |F_1|^2 (\partial_{X_1} \partial_{X_1^*} \Zuu(|X_1|)+ |F_2|^2 (\partial_{X_2} \partial_{X_2^*} \Zuu(|X_2|)  \Big) U^{\dagger}_i U_j  
\nonumber
\eea
Now, in the special case in which
\be
\label{trick}
X_1=M+\theta^2 F \qquad X_2= M-\theta^2 F
\ee
we have that the terms with first derivatives of $\Zuu$ cancel out 
and we are left only with the second derivatives contribution to the soft masses, which are 
two copies of the same expression.
Hence for this model, with the particular choice (\ref{trick}), the induced soft terms are twice the following contributions 
\bea
\label{def_thresold}
&&
\delta m_{U_{i} U_{j}}^2= \frac{1}{256 \pi^4} c_{ij} \lambda^2 d_U  \left(  \lambda^2 d_{\phi}    -2    \sum_{r=1,3} C_r g_r^2 \right) \Lambda^2
\\
&&
\delta m_{Q_{ij}}^2=-\frac{d_U \lambda^2}{256 \pi^4} (y^{\dagger}.c.y)_{ij} ~\Lambda^2 \nonumber  \\
&&
\delta m_{H_u}^2=-\frac{3 d_U \lambda^2}{256 \pi^4}  \Tr (y^{\dagger}.c.y) ~\Lambda^2 \nonumber 
\eea
\noindent
In the special choice (\ref{trick}), we can rotate the messenger fields in a new basis such that the theory (\ref{complete1}) is equivalent to the following model
\be
W=\lambda \sum_{i=1}^3 c_i U_i (\phi_a \phi_b +\phi_c \phi_d)+M (\phi_a \tilde \phi_a +\phi_b \tilde \phi_b + \phi_c \tilde \phi_c +\phi_d \tilde \phi_d) + 
Y   (\phi_a \tilde \phi_b +\phi_b \tilde \phi_a + \phi_c \tilde \phi_d +\phi_d \tilde \phi_c) 
\ee
where we denoted $Y=\theta^2 F$,
which is exactly two copies of the R-symmetric model considered in the paper, see eq (\ref{R-preserving}) and (\ref{deformation}).
We hence conclude that the soft terms induced by the R-symmetric model at leading order in $F/M$ are given by (\ref{def_thresold}).
Moreover we also computed explicitly the one loop corrections (suppressed in $F/M$) to the soft masses in the R-symmetric model, and found that they are equivalent
to the one induced by the non R-symmetric model, i.e. the last term in $\delta m_{U_{i} U_{j}}^2$ in (\ref{def_thresold_nonR}).

Hence the total contribution to the soft terms for the model discussed in the main body of the paper, i.e. (\ref{R-preserving}) and (\ref{deformation}), 
is indeed the one quoted in (\ref{def_thresold_R}).
We stress that no A-terms are generated, and the $|A|^2$ term is not present in the two loop corrections to the soft masses.

\end{document}

%% file: single_top.tex
Our analysis of the single top channel ($i.e.$ $l+\MET + b$) is inspired by previous work of Ref.~\cite{Aad:2014wza}. We include modifications to optimise the analysis for the high luminosity LHC. 
The single top event topology of signal events results from all the processes depicted in Figure \ref{sup1_pairs}.
The pair production of either same or opposite sign $\tilde{u}$ contribute to this final state when one of the squarks decays into a $\tilde{\chi}_0$ and a $u$-jet, and the other decays into $\tilde{\chi}_0$ and a $t$ quark\footnote{We note that mixing between sup and stop can give rise to loop-induced direct top production\cite{Plehn:2009it}. This contribution could potential increase our signal rate, however, after applying the analysis cuts of Eq.~(\ref{eq:cuts-sing}) direct top production is irrelevant.}.
Analogously, the gluino squark associated production generates single top topology when either the gluino or the squark present at least one top in its decay chain.

\begin{figure}
\begin{center}
 \includegraphics[width=.45\textwidth]{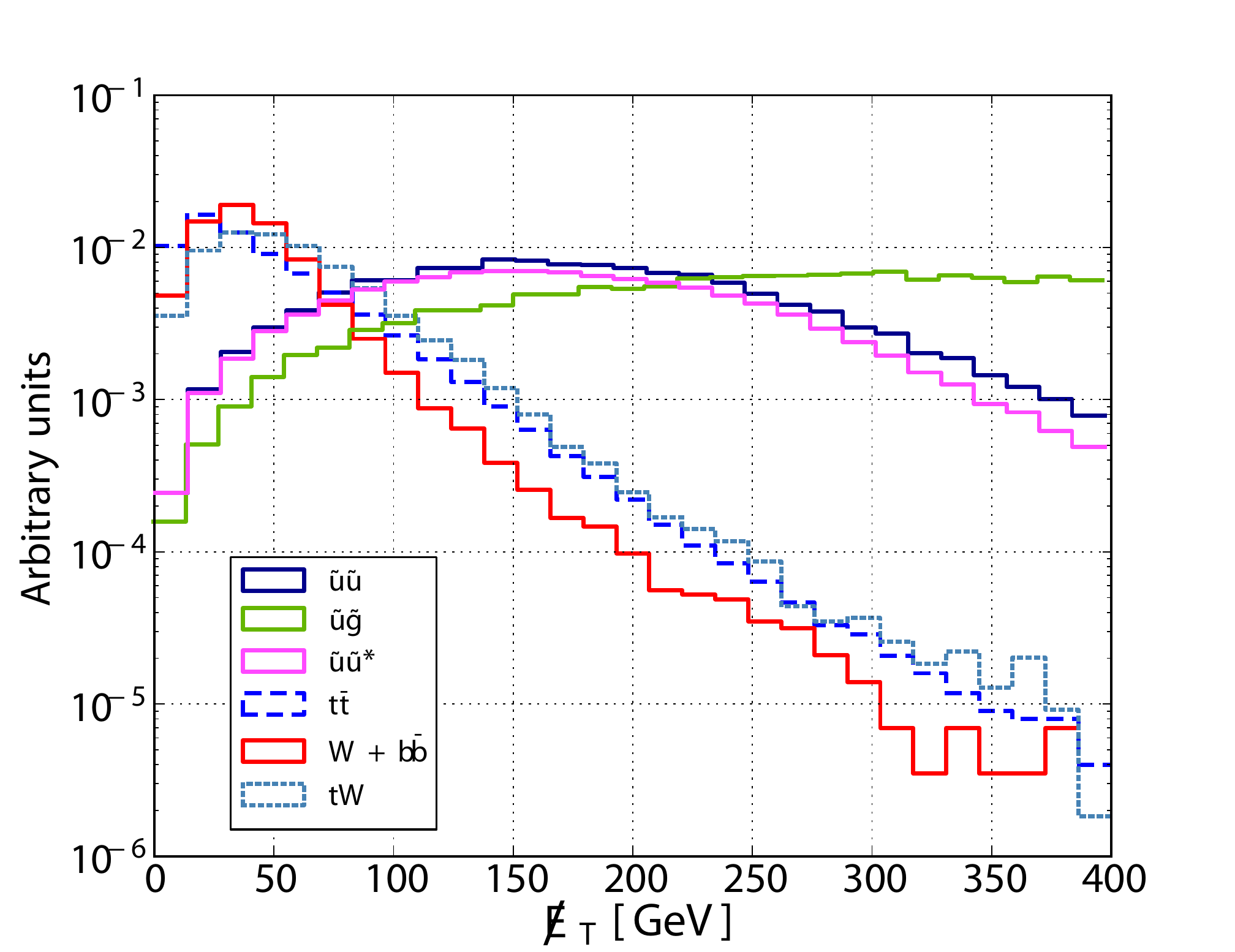}
 \includegraphics[width=.45\textwidth]{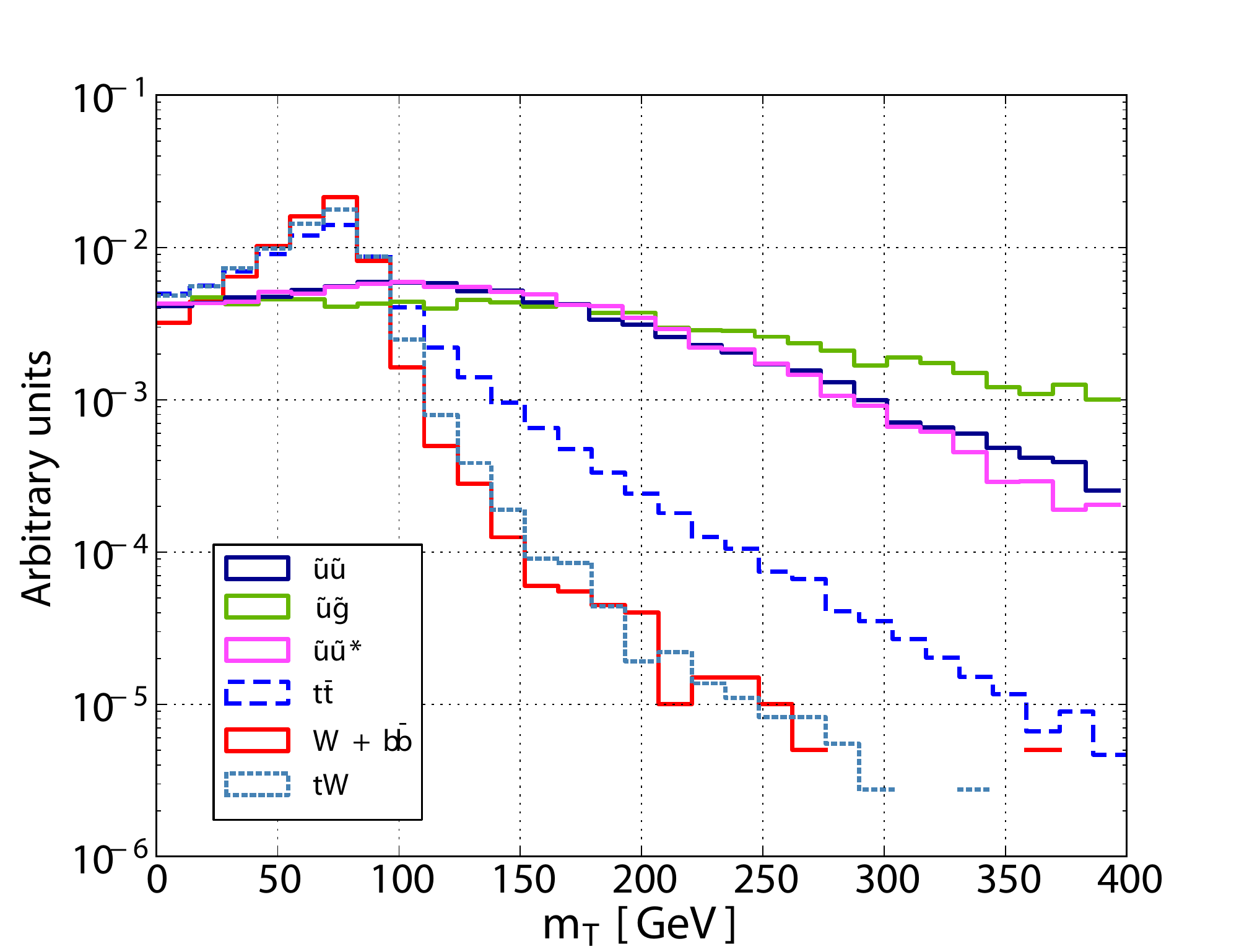}
  \includegraphics[width=.45\textwidth]{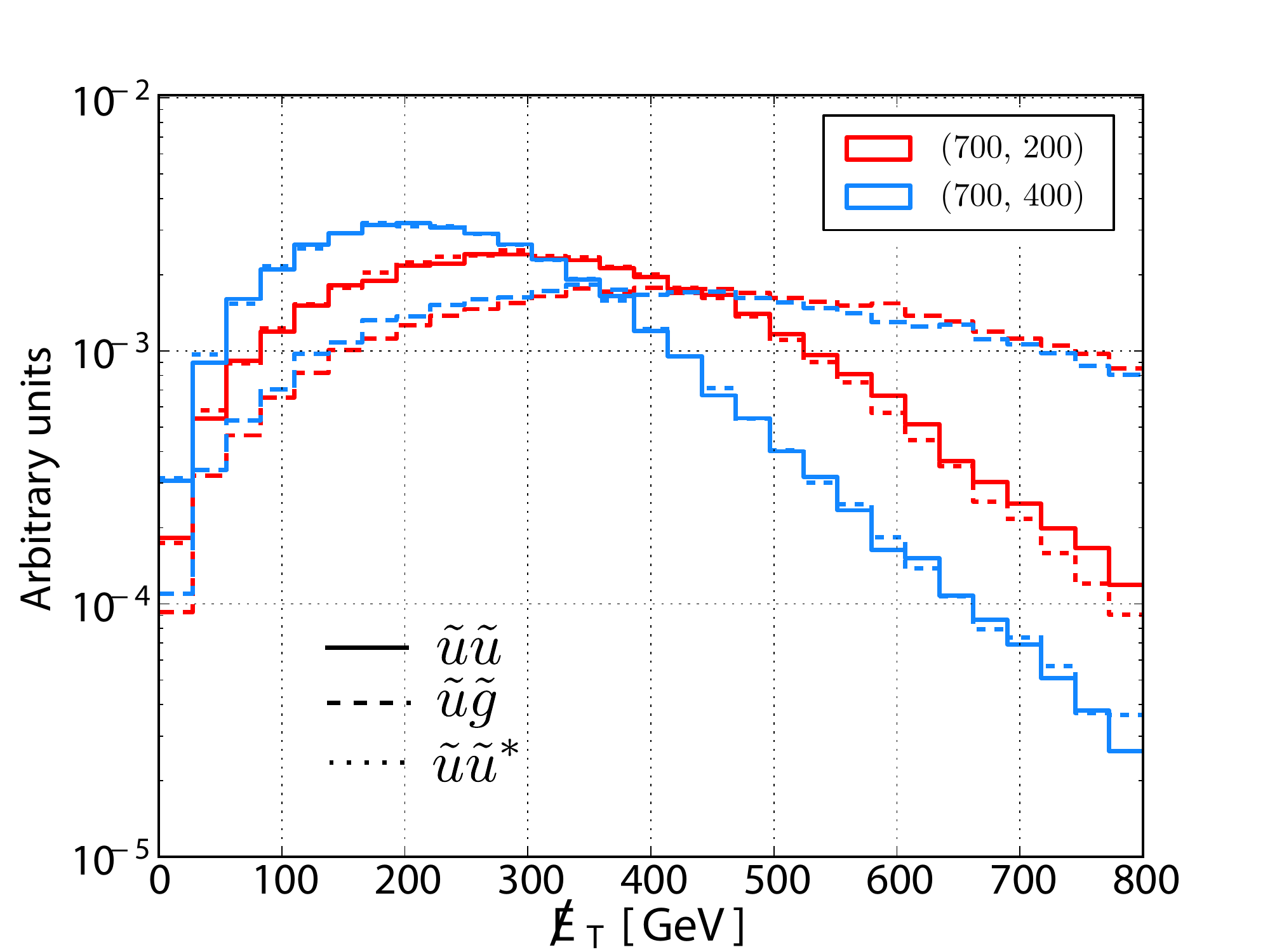}
   \includegraphics[width=.45\textwidth]{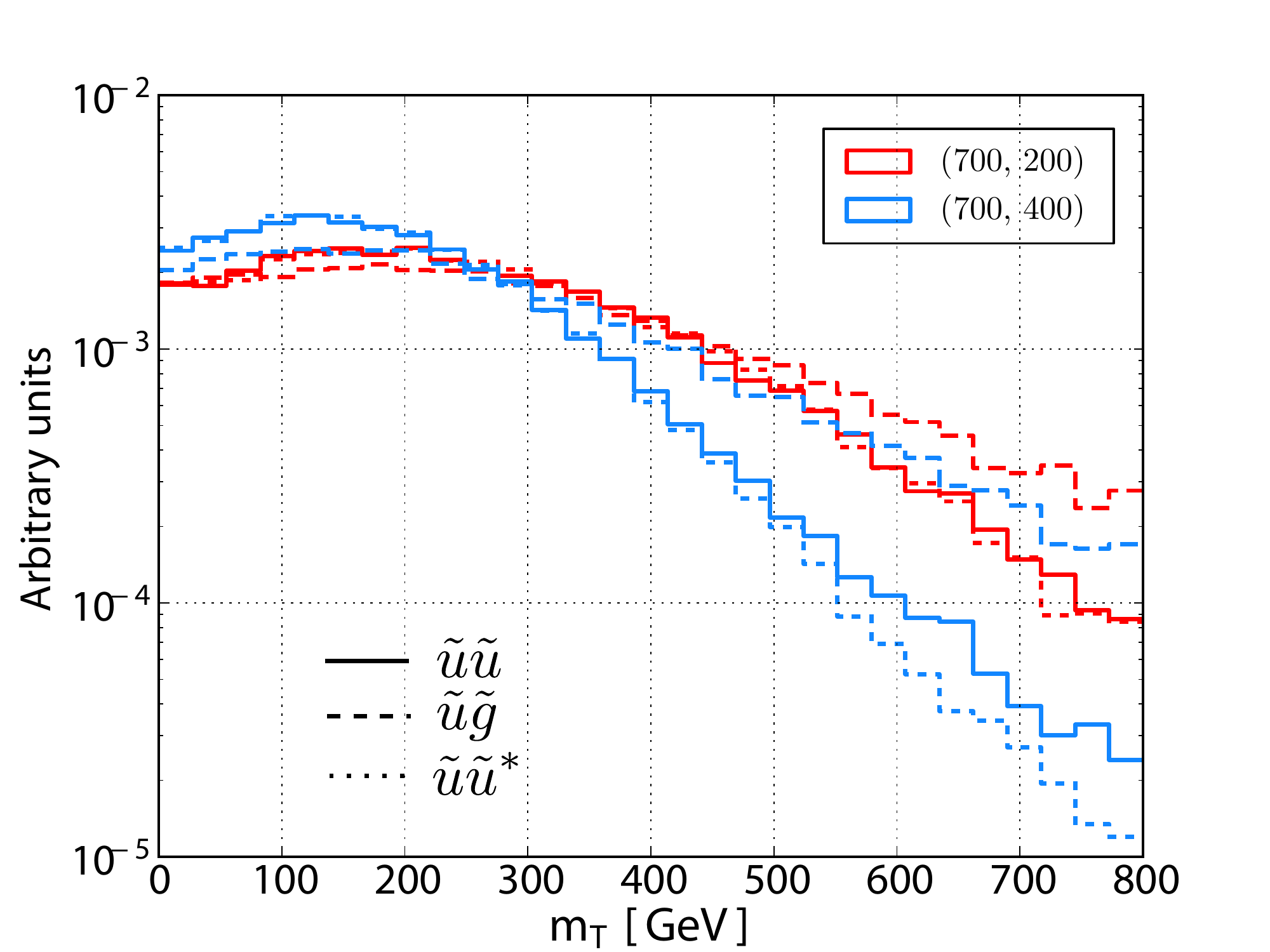}
\caption{Kinematic distributions of signal and background events in the $l + \MET+ b$ channel. All distributions are normalized to unit area and assume no cuts. The signal distribution in the upper panels shows a benchmark point of $m_{\tilde{u}} = 450 \GeV$ and $m_{\tilde{\chi}_0} = 200 \GeV$. In the bottom panels, the values in the labels represent ($m_{\tilde{u}}, m_{\tilde{\chi}_0}$).\label{fig:kinematics_onetop}}
\end{center}
\end{figure}

Fig.~\ref{fig:kinematics_onetop} shows examples of  interesting kinematic distributions for signal and background events. 
Signal events in the single top channel are characterized by large missing energy compared to SM backgrounds, as well as large transverse mass\footnote{Here we define $m_T \equiv \sqrt{2\, p_T^l \, \MET\,\left( 1 - cos(\Delta \phi_{l\, \MET}\right) }. $}. The transverse mass distributions of the SM backgrounds display a suppression around the $W$ mass, as the only source of significant missing energy and hard leptons is the decay of the $W$ boson. Conversely, a significant contribution to missing energy in the signal events comes from $\tilde{\chi}_0$, allowing the transverse mass distribution to extend to much larger values. 

The degree of squark-neutralino mass degeneracy has a large effect on the shape of transverse mass and missing energy distributions. Lower panels of Fig.~\ref{fig:kinematics_onetop} illustrate the point for the benchmark mass of $m_{\tilde{u}} = 700 \GeV$. The benchmark point with $m_{\tilde{\chi}_0} = 400 \GeV, $ leads to much softer $\MET$ and $m_T$ spectra compared to $m_{\tilde{\chi}_0} = 200 \GeV$ in $\tilde{u}\tilde{u}$ and $\tilde{u}\tilde{u}^*$ production, while the effect is much milder in the $\tilde{u}\tilde{g}$ production. The effect suggests that a cut on $m_T$ or $\MET$ which would be appropriate to isolate the signal in the compressed mass region, might not be optimal in the non-compressed spectrum. For the purpose of illustration, here we will focus only on cut selection criteria which can better probe the uncompressed mass spectrum scenario. 

In order to improve the purity of the event sample, we impose the following set of kinematic cuts:
\begin{eqnarray}
\label{eq:cuts-sing}
	N_l (p_T > 20 \GeV, \, \eta < 2.5) = 1\,, \,\,\,\,\,& N_b (p_T > 20 \GeV, \, \eta < 2.5) = 1\, , \nonumber \\
	\MET > 250 \GeV\,, \,\,\,\,\,\,& m_T > m_T^{\mathrm{min}}\,,
\end{eqnarray}
where $b$ and $l$ refer to $b$-tagged jets and isolated leptons respectively, and we choose $ m_T^{\mathrm{min}} = (210, 310, 410) \GeV$ for $m_{\tilde{u}} = 450, 700, 950 \GeV$ respectively. 

Table \ref{tab:cutflow_singletop} shows an example cutflow for a benchmark signal points of $m_{\tilde{u}} = 450, 950$ GeV and $m_{\tilde{\chi}_0} = 200, 400$ GeV. Requiring exactly one lepton and at least one $b$-tagged jet is sufficient to bring the signal to background ratio ($S/B$) to levels of $\sim 10^{-3}$ for lighter squark masses, but only to $\sim 10^{-5}$ if the squark mass is $\sim 1 \TeV$.
A cut on missing energy results in a factor of $\sim 100$ improvement in $S/B$,  while the additional cut on $m_T$ improves $S/B$ by an additional factor of 10 at a $50 \%$  signal loss. Our results, summarised in Table~\ref{tab:STsummary}, show that the $m_{\tilde{u}} = 450 \GeV, \, m_{\tilde{\chi}_0} = 200 \GeV$ benchmark point is discoverable with the signal significance of $S/\sqrt{B}  \gg 5$ with $L = 300 \fb^{-1}$ of integrated luminosity at LHC Run II, while with the same amount of data we should be able to rule out our model for squark masses of roughly $\lesssim 1 \TeV$~\footnote{The cuts we chose for this analysis are somewhat optimised for high luminosity LHC. Relaxing the $m_T$ and $E_T$ cuts at lower integrated luminosities is also likely to be more efficient in signal regions with smaller cross sections. }. 

\begin{table}
\begin{center}
\setlength{\tabcolsep}{.3em}
\begin{tabular}{c|ccc|ccc|c|c}
\multicolumn{9}{c}{$m_{\tilde{u}} = 450 \GeV, m_{\tilde{\chi}_0} = 200 \GeV, m_{\tilde{g}} = 2 \TeV$ }\\
\hline
 $l +\MET + b$ & $\tilde{u} \tilde{u}$ & $\tilde{u}\tilde{g}$ & $\tilde{u} \tilde{u}^*$& \,\,\,\, $t\bar{t}$ \,\,\,\,\, & $W+b\bar{b}$ & $W+t$ & $S/B$ &$S/\sqrt{B}\,(300 \fb^{-1})$ \\
\hline
$N_l = 1, N_b = 1$  & 2.6 & 39.0 & 1.1 & $4.1 \times 10^4 $  & 420.0 & 405.0 &  $1.0 \times 10^{-3}$ & 3.6  \\
$\MET> 250 $ GeV &0.67  & 0.82 & 8.3 &  238.0 &16.0  & 9.3 & $ 0.037 $ & 10.0 \\
$m_T> 210 $ GeV & 0.27 & 0.40 & 3.4 & 12.0 & 0.15 & $< 0.1$ & 0.32 & 20.0 \\
\end{tabular}

\begin{tabular}{c|ccc|ccc|c|c}
\multicolumn{9}{c}{$m_{\tilde{u}} = 950 \GeV, m_{\tilde{\chi}_0} = 400 \GeV, m_{\tilde{g}} = 2 \TeV$ }\\
\hline
 $l + \MET+ b$ & $\tilde{u} \tilde{u}$ & $\tilde{u}\tilde{g}$ & $\tilde{u} \tilde{u}^*$& \,\,\,\, $t\bar{t}$ \,\,\,\,\, & $W+b\bar{b}$ & $W+t$ & $S/B$ &$S/\sqrt{B}\,(300 \fb^{-1})$ \\
\hline
$N_l = 1, N_b = 1$  & 0.37 &  0.36 & 0.48 & $4.1 \times 10^4 $  & 420.0 & 405.0 & $2.9 \times 10^{-5}$ &  0.10 \\
$\MET > 250 $ GeV & 0.26 & 0.30 & 0.34  &  238.0 &16.0  & 9.3 & $3.4 \times 10^{-3}$ & 0.96  \\
$m_T> 410 $ GeV & 0.093 & 0.12 & 0.13  & 3.1 & $<0.1$ & $< 0.1$ & 0.11 & 3.4 \\
\end{tabular}

\end{center}
\caption{Example cutflow in the $l + \MET + b$ channel. The entries show cross sections in fb after each consecutive cut.  $W$+jets and $tW$ channels includes a generation level cut of $\MET > 80 \GeV$ in order to improve the statistics in the signal region.  \label{tab:cutflow_singletop}}
\end{table}

\begin{table}
\begin{center}
\begin{tabular}{c|ccc}
$m_{\tilde{\chi}_0} / m_{\tilde{u}}\, (\GeV\,)$ & 450 & 700 & 950\\
\hline
    200 & (0.32, 20.0)& (0.24, 11.0) & (0.12,3.8)\\
    400 & - & (0.11, 5.3)& (0.11,3.4) 

\end{tabular}

\caption{Summary of reach for benchmark points in the single top channel. The table entries show $S/B$ and $S/\sqrt{B}$ at 300$\fb^{-1}$ respectively which can be achieved at LHC Run II in the single top channel. The masses of squarks and neutralinos are listed on in the topmost row and leftmost column respectively. All results assume $m_{\tilde{g}} = 2 \TeV$. \label{tab:STsummary} }

\end{center}
\end{table}


A potential discovery of a signal in the $l+\MET + b$ channel would give indirect evidence for the existence of supersymmetry, but would not provide information on the degree (if any) of the $\tilde{u}-\tilde{t}$ mixing, as even then minimal flavor-conserving SUSY models predict signals in the single-lepton channel. Measuring additional channels would be required to determine the presence of $\tilde{u}-\tilde{t}$ mixing, of which we find that the channel with two positively charged leptons is an excellent candidate.

%% file: same_sign.tex
Within the framework of SUSY \footnote{Other non-supersymmetric models such as $Z'$, Composite Top and extra-dimensional models can also produce final states with two positive leptons. }, the ``smoking gun'' signal of the maximally mixed $\tilde{u}$ model at the LHC  is the final state containing two positive leptons and large missing energy. 

The same-sign positive lepton final states are a consequence of the $uu \rightarrow \tilde{u}\tilde{u}$ process, with consecutive decays to $t\chi_0$,
or of the $u g \rightarrow \tilde u \tilde g$ process, where $\tilde u$ decay to $t\chi_0$ and the decay chain of the gluino present one positive top.
Though suppressed by two powers of the $\tilde{u} - \tilde{t}$ mixing angle, as well as the small branching ratios of $W$ to leptons, the $l^+l^+ + \MET$ channel offers a very clean probe of the presence of large $\tilde{u} - \tilde{t}$ mixing. 
Production of $\tilde{u}^* \tilde{u}^*$ and $\tilde g \tilde{u}^*$ which would yield two negative sign leptons in the final state, contributes only few $\%$ to the total signal cross section, due to the PDF suppressions. In the context of SUSY, the strong PDF suppression is a valuable feature of signals with large $\tilde{u}\tilde{t}$ mixing, as other RP conserving supersymmetric models can predict same-sign lepton signals with 
the same amounts of $l^+l^+$ and $l^-l^-$ events \cite{Martin:2008aw}. Furthermore typical RPV models 
with Baryon violation lead to signals with dominant $l^-l^-$ \cite{Durieux:2012gj}, while RPV models which also includes lepton number violation
could lead to dominant $l^+l^+$ \cite{Csaki:2013jza}.

\begin{figure}[htb]
\begin{center}
 \includegraphics[width=.45\textwidth]{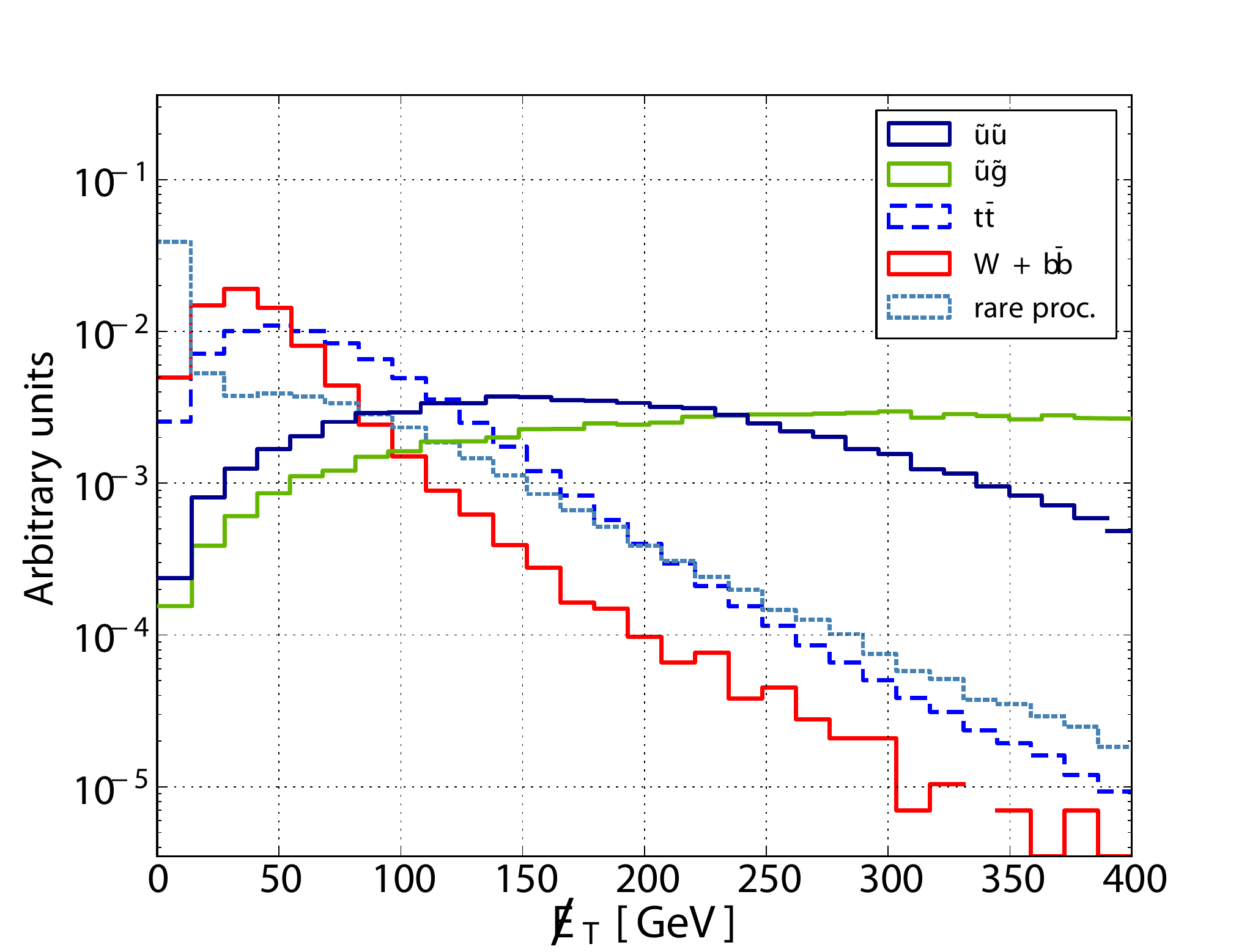}
 \includegraphics[width=.45\textwidth]{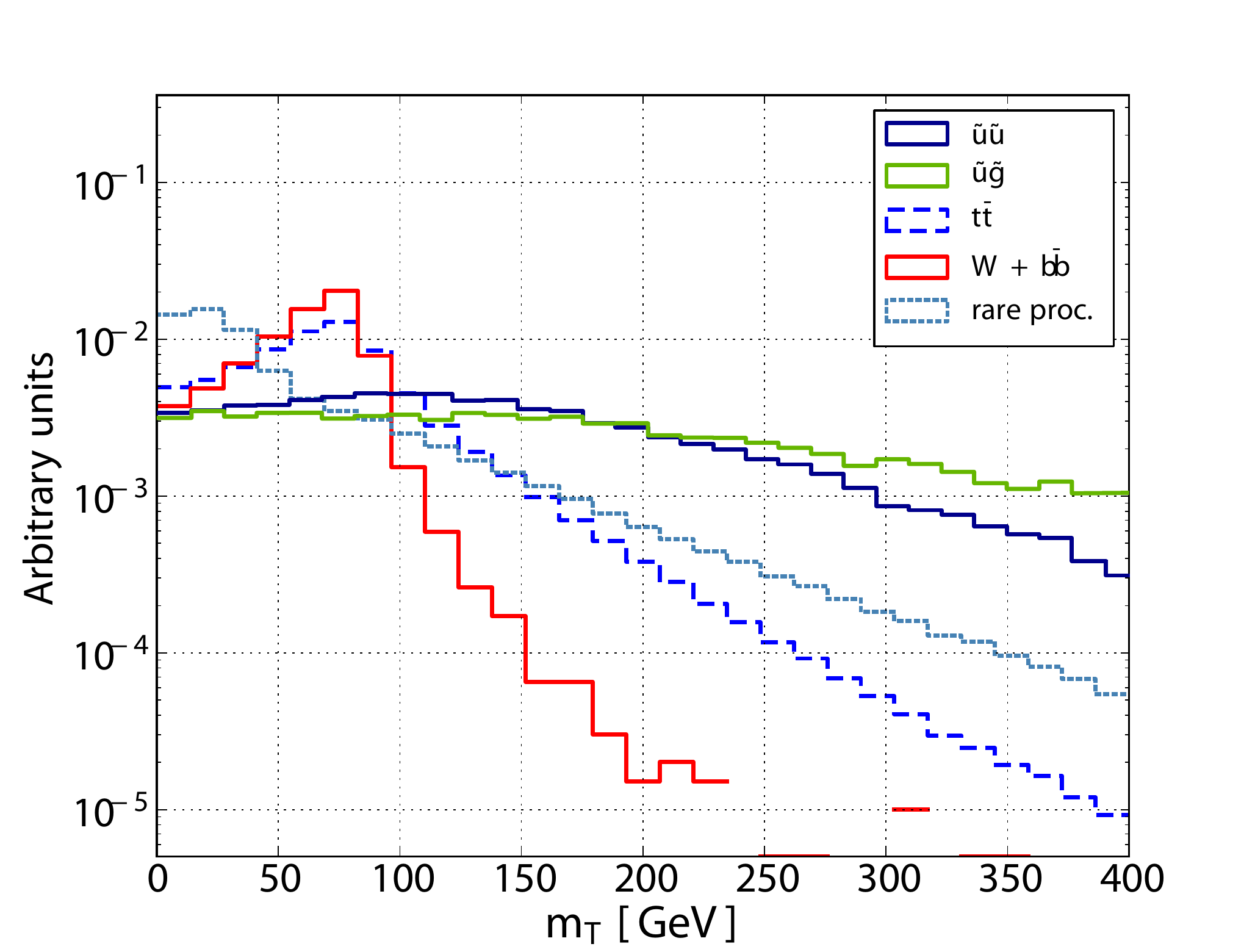}
  \includegraphics[width=.45\textwidth]{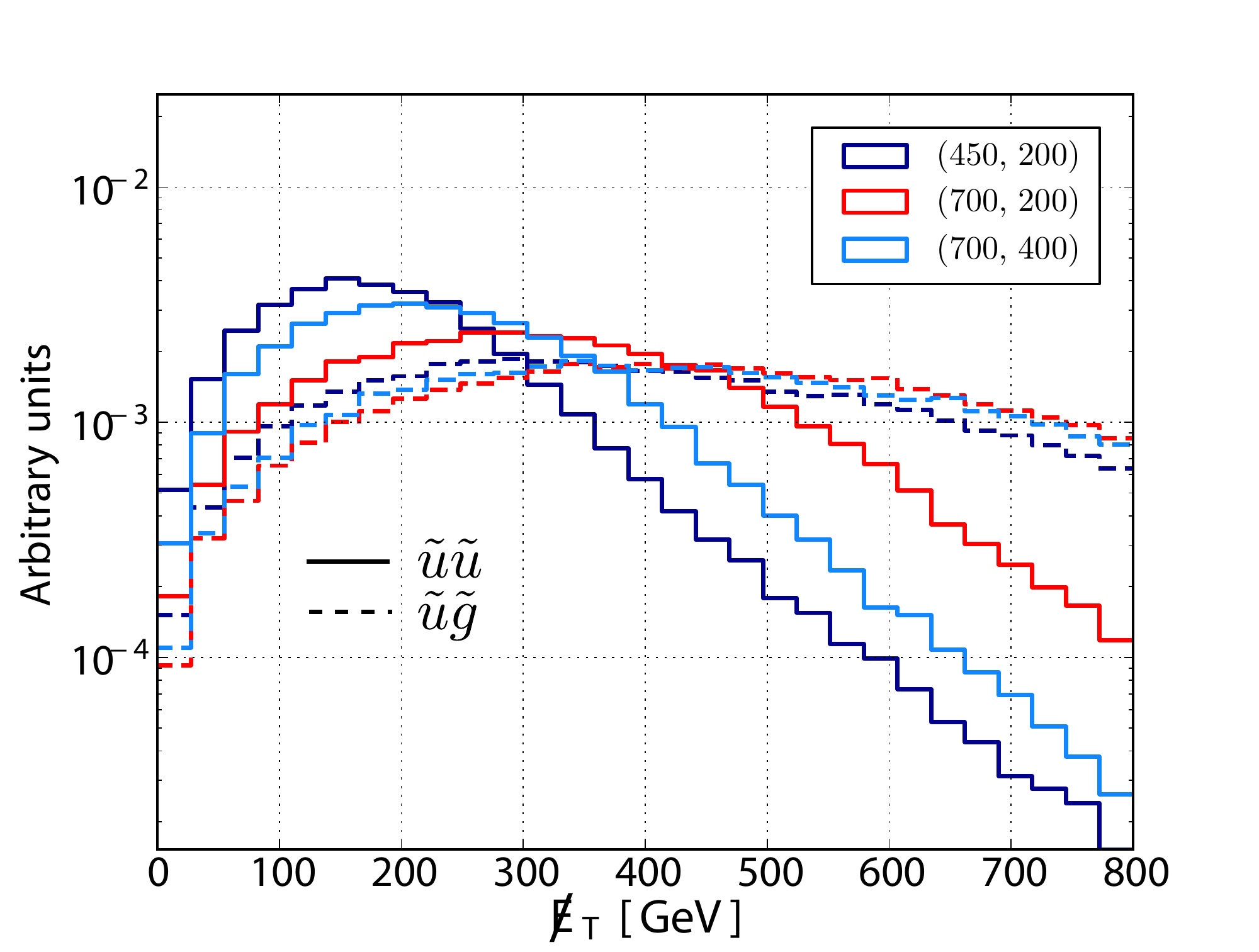}
 \includegraphics[width=.45\textwidth]{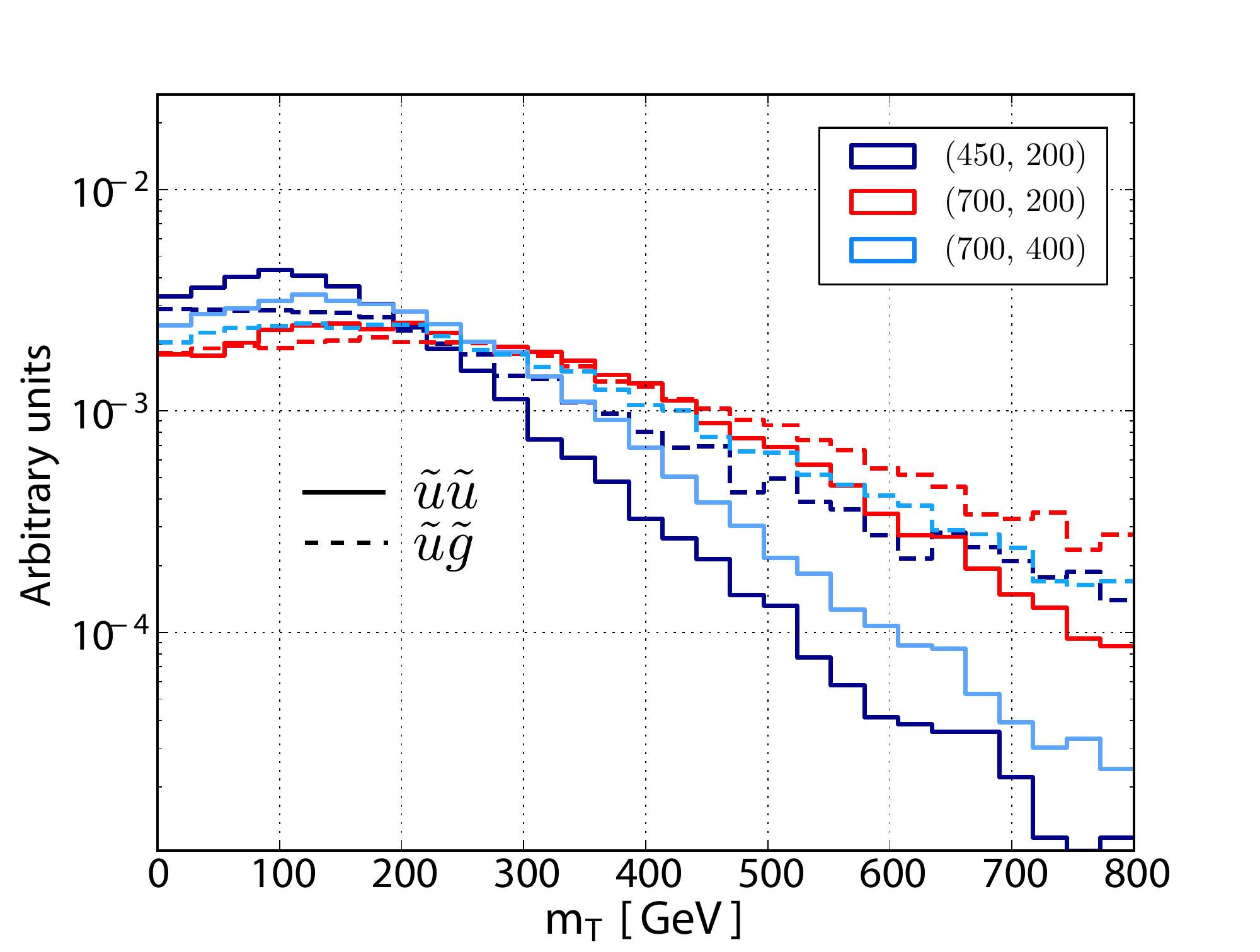}

\end{center}
\caption{Kinematic distributions relevant for the signal search in the $l^+ l^+ + \MET$ channel. All histograms are normalised to unit area. The label ``rare proc.'' includes SM production of $t\bar{t}W, t\bar{t}Z, ZZ, W^+ W^+ W^-$ and $WZ$. In the top panels, we are showing only the benchmark point with $( m_{\tilde{u}}, m_{\tilde{\chi}_0}) = (450, 200) \GeV $.  In the bottom panels, the values in the legend represent different ($ m_{\tilde{u}}, m_{\tilde{\chi}_0}$) benchmark points\label{fig:ssl_kinematics}.}
\end{figure}

Fig.~\ref{fig:ssl_kinematics} shows the characteristic kinematic distributions of the signal and background  in the $l^+ l^+ + \MET$  channel. SM backgrounds consist mainly of $\mathcal{O}(10)$~fb level rare SM processes (here we consider $t\bar{t}W, t\bar{t}Z, ZZ, W^+ W^+ W^-$ and $W^+Z$), as well as SM $t\bar{t}$ and $W+b\bar{b}$, where the same sign lepton background comes mainly from leptonic $b$ decays which yield isolated leptons. With the exception of rare processes, the probability that SM processes contain two positive isolated leptons is tiny, yet significant due to the large production cross sections. In addition, both the amount of missing energy and the transverse mass of the signal events are much larger than in the SM backgrounds, as shown in the right panel of Fig.~\ref{fig:ssl_kinematics}.

In order to maximise the signal significance in the same-sign positive lepton channel, here we employ a set of cuts:
\begin{eqnarray}
	N_{l^+} (p_T > 20 \GeV, \, \eta < 2.5) = 2\, ,   \,\,\,\,\,\,& \MET> 120 \GeV\,, \nonumber  \\
         N_j(p_T > 30 \GeV, \, \eta < 2.5) > 1\,, \,\,\,\,\, &	m_T(l^+_1) > m_T^{\mathrm{min}} \,, \label{eq:cutsSSL} \nonumber \\
         N_{l^{-}} (p_T > 20 \GeV,\, \eta < 2.5) =0 \,\,\,\,\,\,&
\end{eqnarray}
where $l^{+-}$ refer to isolated positive/negative leptons and $l^+_{1}$ refers to the highest $p_T$  positive lepton in the event. For the purpose of illustration,  we set $m_T^{\mathrm{min}}  =(200, 300, 400) \GeV$, for $m_{\tilde{u}} = (450, 700, 950) \GeV$ respectively. 

Our choice of cuts on $\MET$ and $m_T$ was somewhat optimised to probe high $m_{\tilde{u}}$ in a non-compressed mass spectrum scenario. Again, it is important to note that for the purpose of probing the compressed region of $ m_{\tilde{u}}, m_{\tilde{\chi}_0}$ parameter space it is beneficial to relax the cuts on missing energy and transverse mass. Bottom panels of Figure~\ref{fig:ssl_kinematics} illustrate the point. The benchmark point of $ (m_{\tilde{u}}, m_{\tilde{\chi}_0}) = (700, 400) \GeV$ displays a much softer spectrum of missing energy and transverse mass than the corresponding, non-compressed point of $ (m_{\tilde{u}}, m_{\tilde{\chi}_0}) = (700, 200) \GeV$, suggesting that a cut on $\MET$ and $m_T$ lower than the one we suggest in Eq.~\ref{eq:cutsSSL} could be more optimal in the compressed spectrum scenarios. 

Continuing, Table \ref{tab:cutflow_SSL} shows an example cutflow for two benchmark parameter points. The requirement on the lepton multiplicity efficiently reduces the $t\bar{t}$ and $W + b\bar{b}$ backgrounds, while the $\MET$ cut efficiently suppresses the rare-process contribution to the total event yield. We find that a minimal set of cuts in Eq.~\ref{eq:cutsSSL} results in a factor of $\sim 10 -15$ improvement in $S/B$, at a cost of $\sim 50 \%$ in signal efficiency.

\begin{table}[htb]
\begin{center}
\setlength{\tabcolsep}{.3em}
\begin{tabular}{l|cc|ccc|c|c}
\multicolumn{8}{c}{$m_{\tilde{u}} = 450 \GeV, m_{\tilde{\chi}_0} = 200 \GeV, m_{\tilde{g}} = 2 \TeV$ }\\
\hline
 $l^+l^+ +\MET+jj$& $\tilde{u} \tilde{u}  $& $\tilde{u} \tilde{g}$& \,\,\,\, $t\bar{t}$ \,\,\,\,\, & $W$+jets & rare proc. & $S/B$ & $S/\sqrt{B}\,(3000 \fb^{-1})$ \\
\hline
$N_{l^+} = 2, N_{l^-} =0$ & 0.21 & 0.037  &  0.067 & 0.022 &19.0 & 0.013  & 3.1 \\
$N_j > 1$ &  0.20 & 0.037 & 0.033 & 0.022 & 14.0 & 0.017  & 3.4  \\
$\MET > 120$ GeV & 0.12 & 0.034 & 0.033 & $<0.01$  & 1.2  & 0.12 & 7.6\\
$m_T > 200$ GeV & 0.080 & 0.019  &$< 0.01$ & $< 0.01$ & 0.45 &  0.21 & 7.9
\end{tabular}

\begin{tabular}{l|cc|ccc|c|c}
\multicolumn{8}{c}{$m_{\tilde{u}} = 950 \GeV, m_{\tilde{\chi}_0} = 400 \GeV, m_{\tilde{g}} = 2 \TeV$ }\\
\hline
 $l^+l^+ + \MET+jj$& $\tilde{u} \tilde{u}  $& $\tilde{u} \tilde{g}$& \,\,\,\, $t\bar{t}$ \,\,\,\,\, & $W$+jets & rare proc. & $S/B$ & $S/\sqrt{B}\,(3000 \fb^{-1})$ \\
\hline
$N_{l^+} = 2, N_{l^-} =0$ & 0.031 & 0.014  &  0.067 & 0.022 &19.0 & 0.0023 & 0.56 \\
$N_j > 1$ &  0.030 & 0.014 & 0.033 & 0.022 & 14.0 &  0.0031 & 0.64 \\
$\MET > 120$ GeV & 0.026 & 0.014 & 0.033 & $<0.01$  & 1.2  & 0.033 & 2.0 \\
$m_T > 400$ GeV & 0.014  &0.0068 &$< 0.01$ & $< 0.01$ & 0.081 & 0.26  & 4.0
\end{tabular}

\end{center}
\caption{Example cutflow in the $l ^+l^+ + \MET $ channel. The entries show cross sections in fb after each consecutive cut. The label ``rare proc.'' includes SM production of $t\bar{t}W, t\bar{t}Z, ZZ, W^+ W^+ W^-$ and $WZ$. We compute $m_T$ using the hardest positive lepton in the event and $\MET$.  \label{tab:cutflow_SSL}  }
\end{table}

Table \ref{tab:SSLsummary} shows a summary of results on five benchmark points in the $m_{\tilde{u}}, m_{\tilde{\chi}_0}$ space. We find that the LHC can achieve a signal significance of $5 \sigma$ for $\tilde{u}$ with masses of up to $\sim 700$ GeV for a neutralino mass $\lesssim 400 \GeV$. A significance higher than $3 \sigma$ can be achieved for masses up to $\sim 1 \TeV$, assuming the neutralino masses of $\lesssim 400 \GeV$, suggesting that the same sign positive lepton channel could rule out the model up to $\sim 1 \TeV$ at the high luminosity LHC.

\begin{table}
\begin{center}
\begin{tabular}{c|ccc}
$m_{\tilde{\chi}_0} / m_{\tilde{u}}\, (\GeV\,)$ & 450 & 700 & 950\\
\hline
    200 & (0.21, \, 7.9)& (0.31, 7.5) & (0.26,  4.1)\\
    400 & - & (0.22 , 5.3)& (0.26, 4.0) 

\end{tabular}

\caption{Summary of reach for benchmark points in the same sign positive lepton channel. The table entries show $S/B$ and $S/\sqrt{B}$ at 3000 $\fb^{-1}$ respectively which can be achieved at LHC Run II in the same-sign positive lepton channel. The masses of squarks and neutralinos are listed on in the topmost row and leftmost column respectively. All results assume $m_{\tilde{g}} = 2 \TeV$. \label{tab:SSLsummary} }

\end{center}
\end{table}

The reach of the single top channel at $300 \fb^{-1}$ is comparable to the reach of the same-sign positive leptons search at $3000 \fb^{-1}$. In case a signal is observed in the single top channel at $300 \fb^{-1}$, the high luminosity LHC should be able to probe and measure possible large squark mixings.  Conversely, in case no signal is observed in the single top channel at $300 \fb^{-1}$, the parameter region giving raise to measurable same-sign positive leptons yield at High Luminosity LHC will already be ruled out.